\documentclass[]{servicenow} 

\usepackage{amsmath,amsfonts,bm}

\def\eqref#1{equation~\ref{#1}}

\def\1{\bm{1}}

\DeclareMathAlphabet{\mathsfit}{\encodingdefault}{\sfdefault}{m}{sl}
\SetMathAlphabet{\mathsfit}{bold}{\encodingdefault}{\sfdefault}{bx}{n}

\usepackage{amsmath,amsfonts,bm}

\def\eqref#1{equation~\ref{#1}}

\def\1{\bm{1}}

\DeclareMathAlphabet{\mathsfit}{\encodingdefault}{\sfdefault}{m}{sl}
\SetMathAlphabet{\mathsfit}{bold}{\encodingdefault}{\sfdefault}{bx}{n}

\usepackage[textsize=tiny]{todonotes}
\usepackage{lmodern}
\usepackage{natbib}
\usepackage[utf8]{inputenc} 
\usepackage[T1]{fontenc}   
\usepackage{amsfonts}  
\usepackage{nicefrac}  
\usepackage{subcaption}
\usepackage{enumitem}
\usepackage{wrapfig}  
\usepackage{amsmath}
\usepackage{amssymb}
\usepackage{mathtools}
\usepackage{amsthm}
\usepackage{hyperref}
\usepackage{url}
\usepackage{microtype}
\usepackage{graphicx}
\usepackage{booktabs}
\usepackage{newtxtext} 
\usepackage{breakurl}
\usepackage{adjustbox}
\usepackage{float}

\newcommand{\ourmodel}{{\textsc{ReTreever}}}
\newcommand{\B}[1]{\mathbf{#1}}

\newcommand{\cP}{\ensuremath \mathcal{P}}

\newcommand{\eb}{\ensuremath \B{e}}
\newcommand{\xb}{\ensuremath \B{x}}

\newcommand{\cb}{\ensuremath \B{c}}

\newcommand{\zb}{\ensuremath \B{z}}
\newcommand{\qb}{\ensuremath \B{q}}

\DeclareMathOperator*{\simi}{sim}

\title{Hierarchical Retrieval at Scale: Bridging Transparency and Efficiency}

\author[1,3,4]{Shubham Gupta}
\author[1,2,4]{Zichao Li}
\author[1]{\\Tianyi Chen}
\author[3,4]{Cem Subakan}
\author[1,2,4]{Siva Reddy}
\author[1,2,4]{Perouz Taslakian}
\author[1,3]{Valentina Zantedeschi}

\affiliation[1]{ServiceNow Research}
\affiliation[2]{McGill University}
\affiliation[3]{Universit\'e Laval}
\affiliation[4]{Mila -- Qu\'ebec Artificial Intelligence Institute}

\abstract{
Information retrieval is a core component of many intelligent systems as it enables conditioning of outputs on new and large-scale datasets. While effective, the standard practice of encoding data into high-dimensional representations for similarity search entails large memory and compute footprints, and also makes it hard to inspect the inner workings of the system. Hierarchical retrieval methods offer an interpretable alternative by organizing data at multiple granular levels, yet do not match the efficiency and performance of flat retrieval approaches. In this paper, we propose \ourmodel{}, a tree-based method that makes hierarchical retrieval viable at scale by directly optimizing its structure for retrieval performance while naturally providing transparency through meaningful semantic groupings. Our method offers the flexibility to balance cost and utility by indexing data using representations from \emph{any} tree level. We show that \ourmodel{} delivers strong coarse (intermediate levels) and fine representations (terminal level), while achieving the highest retrieval accuracy at the lowest latency among hierarchical methods. These results demonstrate that this family of techniques is viable in practical applications.
}

\correspondence{\email{shgup1@ulaval.ca}}

\begin{document}

\maketitle

\section{Introduction}
\label{sec:intro}

Modern intelligent systems increasingly rely on retrieval mechanisms to condition their outputs on new and large-scale datasets. 
In particular, retrieval enables Large Language Models (LLMs) to generate content grounded in external knowledge, thereby alleviating hallucinations \citep{shuster2021-retrieval-augmentation},  while scaling to massive datasets without increasing the number of internal parameters.
As such, retrievers have become a key component in transforming LLMs into tailored experts without the need for exhaustive fine-tuning \citep{lewis2020retrieval,izacard-grave-2021-leveraging,gao2023retrieval}.
A popular retrieval approach consists of representing items as high-dimensional embeddings and use nearest-neighbor techniques to find relevant content for a given query embedding~\citep{karpukhin-2020-dpr}. 
While effective, the high dimensionality of these embeddings results in a large memory footprint and heavy computation at inference time. 
The items are also stored without any human-readable structure or understandable grouping, making it difficult to derive insights from the data or verify the retrieval process.

\begin{figure*}
  \centering
  \begin{subfigure}[t]{0.42\textwidth}
    \centering
    \includegraphics[width=\textwidth]{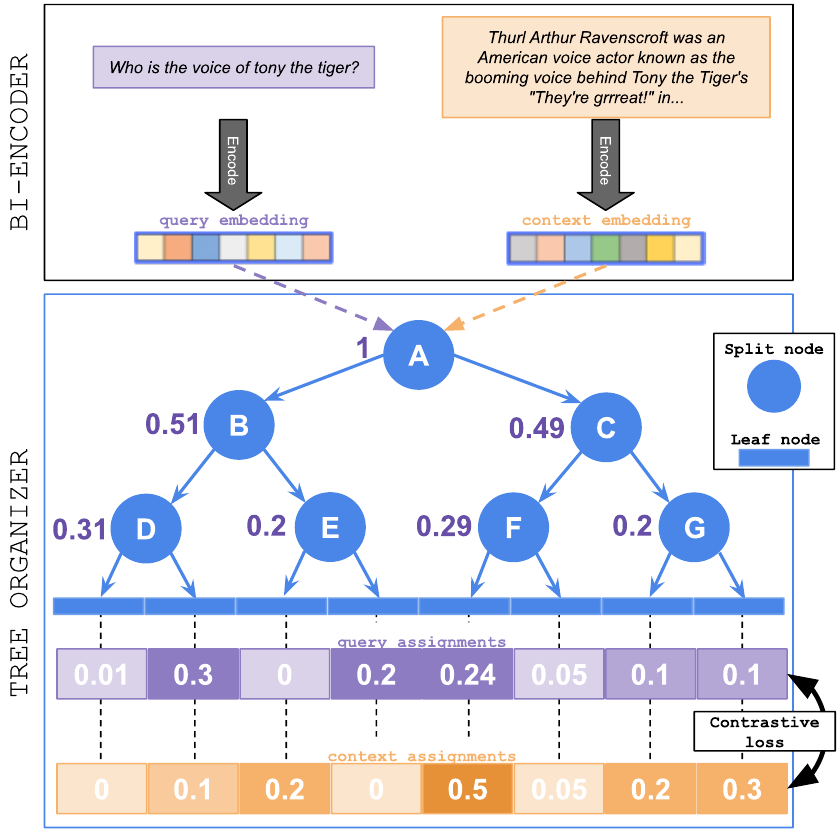}
    \caption{\ourmodel{} architecture and traversal}
    \label{fig:architecture}
  \end{subfigure}
  \hspace{0.05\textwidth}
  \begin{subfigure}[t]{0.47\textwidth}
    \centering
    \includegraphics[width=\textwidth]{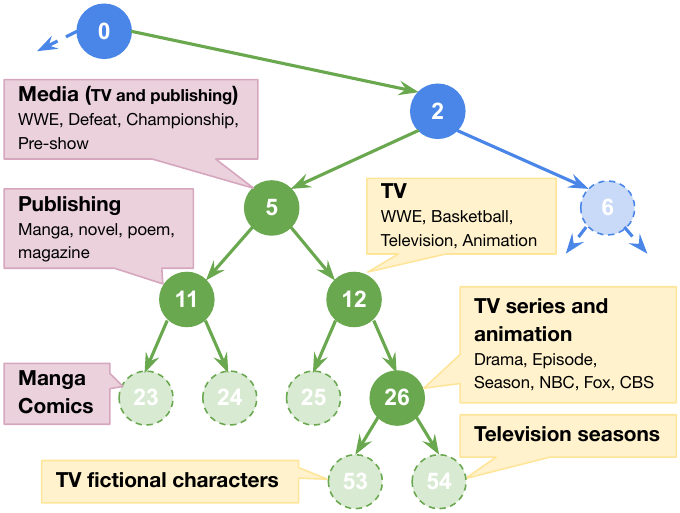}
    \caption{Inspecting a \ourmodel{} tree learned on \textsc{NQ}}
    \label{fig:transparency}
  \end{subfigure}
  \caption{\textbf{(a)} During training, a positive query–context pair is encoded by the frozen Bi-Encoder~$E$; these embeddings pass through each split node of a depth-$D$ tree to yield left/right routing probabilities, which are merged into a leaf-assignment vector (one probability per leaf). At inference, leaf-level probabilities provide fine representations and intermediate-node probabilities serve as coarser representations. \textbf{(b)} Visualization of the topics (bold) and keywords extracted from the contexts assigned to one subtree (in green) rooted at node 5 of a \ourmodel{} tree of depth $10$ trained on \textsc{NQ}. For compactness, we represent only a subset of the nodes,
and stop at depth 5. Topics are locally coherent along a path, indicating that \ourmodel{} naturally groups contexts semantically.}
  \label{fig:tree-combined}
  \vspace{-15pt}
\end{figure*}

In this paper, we propose \ourmodel{}, a tree-based method where documents are organized and represented at various granular levels, offering coarse-to-fine representations, learned entirely end-to-end by optimizing for retrieval performance as the learning objective. 
These representations offer the flexibility to balance cost and utility.
Rather than training separate models for each size, a single full-capacity tree is learned, and at inference one can simply choose a tree level, higher for richer longer embeddings, and lower for smaller coarser ones, so content can be encoded at different sizes to match available computational budget.
\ourmodel{} takes inspiration from the hierarchical retrieval literature, where tree organizers are built and navigated via calls to an LLM by cross-encoding textual queries and contexts~\citep{chen2023walking,sarthi2024raptor,edge2024local}. 
Such retrievers have the advantage of preserving the existing organization of the data, such as document divisions and entity relationships, and grouping documents semantically into a readable tree organization.
However, their reliance on LLMs makes them slow and expensive to train and run, hence impractical even for medium-size text corpora.
In contrast, \ourmodel{} is designed to operate entirely on embeddings, eliminating the need for LLM calls during both construction and navigation.

Figure~\ref{fig:architecture} shows a schematic of our approach for document retrieval. 
\ourmodel{} first converts items (i.e., queries and references) into an embedding using a standard encoder, such as BERT ~\citep{devlin2019-bert} for text and DINOv2 ~\citep{oquab2023dinov2} for images.
It then uses these representations to learn a binary tree structure end-to-end  by optimizing directly the retrieval objective, making use of the query-reference supervision available from most datasets. 
This is achieved by learning the parameters of a routing function at each internal node of the tree, such that positive query-reference pairs are routed similarly through it. 

The learned hierarchical structure naturally provides an organization of the items, which allows us to probe the tree to gain insights into the database and retrieval operations. Practitioners can inspect different levels of the tree to identify key thematic clusters influencing retrieval or understand why certain branches lead to specific content.
While this is not a complete causal explanation of content matching, it does ease the inspection of the model's inner workings.
We provide an interactive tool for inspecting the learned hierarchical organization of a few of our models.\footnote{\label{fn:anon_tool}\url{https://euphonious-brigadeiros-cf5172.netlify.app/}}

In \Cref{sec:interpretability}, we analyze nodes at various levels of the hierarchy and show that documents in these nodes have thematic overlaps even though the tree is solely optimized for query-document similarity without any clustering objective. Our contributions are as follows:

\begin{enumerate}[leftmargin=.6cm, itemsep=-1.5pt]
\item We propose \ourmodel{}, a hierarchical and inspectable retriever that learns coarse-to-fine representations aligned with levels of a retrieval-optimized tree. Unlike existing hierarchical retrievers, \ourmodel{} scales to large corpora without requiring LLM calls.
\item We show that \ourmodel{} achieves the best latency and performance trade-off w.r.t both hierarchical tree-based and representation-based retrievers. 
Such results are consistent across text, image and speech datasets, demonstrating that our framework is modality-agnostic.
\item We highlight how \ourmodel{}’s learned hierarchy induces semantically coherent groupings that can be directly inspected by humans, providing an interpretable window into the retrieval process and supporting better downstream decisions.
\end{enumerate}

\section{Related Work}
\label{app:related work}

\noindent\textbf{Bi-encoders}
Retrieval models typically rely on sparse or dense vector embeddings to select the most relevant items for a query, reducing the problem to a nearest-neighbor search.
Dense retrieval models~\citep{karpukhin-2020-dpr, khattab2020colbert} rely on encoders, neural networks designed to embed the semantic meaning of text into dense vector representations.
State-of-the-art sentence embedding models often leverage bidirectional encoders, such as BERT~\citep{devlin2019-bert} and BGE~\citep{xiao-bge}, 
while others are built on pretrained large language models, such as LLM2Vec~\citep{behnamghader2024llmvec}. 
Some encoders, such as BGE-large~\citep{xiao-bge},
are specifically designed for retrieval, offering a 
sentence embedding framework optimized for such tasks.
Other notable examples are: DPR~\citep{karpukhin-2020-dpr}, which finetunes a BERT dual encoder to separately encode queries and documents into a single dense vector by contrastive learning;
ColBERT~\citep{khattab2020colbert}, 
which finetunes BERT to encode queries and documents into multiple dense vector representations and by matching individual token embeddings between queries and documents.
Unlike sparse retrieval methods, such as BM25 or TF-IDF~\citep{salton-tfidf}, these dense representations 
are less interpretable, expensive to compute (generally because of attention operations) and take up a large amount of storage space. They do however perform better on many downstream tasks such as QA, specially for complex queries. 

\noindent\textbf{Vision and audio encoders} 
Beyond text, dense representation learning has been widely studied for visual and auditory modalities.
In vision, convolutional architectures such as ResNet-50~\citep{he2016resnet} have long served as strong baselines for image embeddings, while transformer-based models such as the Vision Transformer (ViT)~\citep{dosovitskiy2021vit} have become a dominant architecture for learning global visual representations.
Self-supervised approaches built on ViT, such as DINOv2~\citep{oquab2023dinov2}, produce high-quality image embeddings that transfer well across a wide range of downstream tasks without task-specific supervision.
Multimodal models such as CLIP~\citep{radford2021clip} further extend ViT-based encoders by jointly embedding images and text into a shared representation space, enabling cross-modal retrieval.
For audio, transformer-based encoders have similarly emerged as the standard approach.
Audio Spectrogram Transformers (AST)~\citep{gong21b_interspeech} adapt the ViT architecture to time-frequency representations and achieve strong performance on large-scale benchmarks. 
As with text and vision encoders, these models provide compact representations suitable for similarity search, but incur significant costs and offer limited interpretability.

\noindent\textbf{Hierarchical retrieval}
Hierarchical retrieval approaches are designed for text retrieval, and aim to balance test-time efficiency and accuracy by organizing the retrieval process into multiple stages, typically via coarse-to-fine filtering of candidate documents.
Many hierarchical methods face challenges related to computational cost and performance trade-offs, and scale poorly with 
the size of the corpus. 
MemWalker~\citep{chen2023walking} segments long texts into chunks, summarizes them into nodes that are recursively merged into a hierarchical memory tree, and at query time traverses this tree via iterative prompting to stay within the LLM’s context window.
Raptor~\citep{sarthi2024raptor} uses a clustering algorithm to group similar documents and then applies an LLM to recursively summarize and re-embed chunks of text, constructing a tree with differing levels of summarization from the bottom up, resulting in a structured, multi-layered tree representation of the original documents. 
GraphRAG~\citep{edge2024local} uses an LLM to build a graph-based text index by first deriving an entity knowledge graph from source documents and then pregenerating community summaries for related entities. For a given question, partial responses are generated from each community summary and combined into a final summarized response.
While hierarchical retrieval methods improve response quality using LLMs and provides inspectable structures, they incur high computational costs and slow processing times, especially with large datasets. This trade-off makes them less suitable for real-time or resource-limited scenarios, emphasizing the need for more efficient solutions.
\ourmodel{} overcomes these limitations by constructing and navigating a tree with no LLM calls. 

\noindent\textbf{Coarse-to-fine representations}
As compute and running time generally scale with the representation size, an effective solution to limit retrieval costs is through dimensionality reduction\citep{van2009dimensionality}.
When the computational budget is not known in advance, the standard solution is to train multiple models or low-dimensional adapters not to incur into accuracy degradation, as one would by applying post-hoc compression techniques. However, this solution requires higher training and memory costs than learning and storing a single model.
Single-model alternatives have been recently developed~\citep{yu2018slimmable,cai2019once,kusupati2022matryoshka}.
In particular, \citep{kusupati2022matryoshka} propose Matryoshka Representation Learning (MRL), a simple framework that learns a nested representation. 
MRL boils down to learning an adaptive capacity embedding, ensuring that any first m-dimensions vector is as accurate as an independently trained m-dimensional one.
This approach improves retrieval efficiency and flexibility, making it well-suited for diverse and evolving retrieval scenarios.
Similarly, \ourmodel{} benefits from such advantages by training a nested representation, where each level input-to-node assignment corresponds to an embedding. 
Its structure and the learning of assignments additionally provide an organization of the documents into semantic groupings, allowing practitioners to inspect the corpus content and the inner workings of the retrieval system.

\noindent\textbf{Differentiable trees and hierarchical indexes}
Because of their discrete form, tree structures have been traditionally optimized by greedy algorithms, specialized for a particular objective, e.g. for classification, regression, or hierarchical clustering\citep{quinlan1986induction,krishnamurthy2012efficient,quinlan2014c4,breiman2017classification,moseley2023approximation}.
Recent literature has focused on differentiable formulations to take advantage of continuous optimization techniques for training trees on large datasets and for arbitrary objectives\citep{irsoy2012soft,yang2018deep,monath2019gradient,tanno2019adaptive,pmlr-v139-zantedeschi21a,marton2024gradtree}.
We leverage this literature and extend it to learning trees that are optimal for the retrieval objective (via contrastive learning), and with complex split and propagation functions.
Finally, trees and graphs have been used in the retrieval literature as indices for storing and quickly retrieving document embeddings via approximate similarity search, and not designed for being inspected\citep{bernhardsson2017annoy,malkov2018efficient,douze2024faiss}.
\ourmodel{} does not belong to this line of works, as it is an embedding extractor that learns an inspectable tree to organize and represent documents at different granularity and into semantic groupings.

\section{Proposed Method}
\label{sec:proposed_method}
\ourmodel{} consists of (1) a frozen encoder $E$ that produces embeddings for any text input, and (2) a learnable binary tree $T$ that serves as a shared router for both queries and references, organizing references into a hierarchy and routing queries to the relevant region of the tree (see \Cref{fig:architecture}). In this section, we describe how the tree hierarchy is learned by continuous optimization and how search is performed at inference. 
In what follows, we use the term \emph{context} to refer to the content organized by \ourmodel{}, which could be images, reference documents or audio recordings.

\subsection{\ourmodel{} Architecture}
\label{sec:ret_arch}
\paragraph{Tree Formulation.} 

Let $\{(\qb_i, \cb_i) \in \cP \}_{i=1}^n$ be positive query-context pairs  where $\qb_i$, $\cb_i$ are respectively the query and context embeddings generated by the encoder $E$.
Our goal is to learn a binary tree $T$ that assigns $\qb_i$ and~$\cb_i$ to the same leaf node, so that retrieval accuracy is maximized. 
Such leaf assignments, denoted $T(\xb_i)$, are obtained by routing a given input $\xb_i \in \mathcal{X}$ (e.g., $\xb_i \vcentcolon= \qb_i$ the query embedding) sequentially through branching nodes until it reaches the terminal level of the tree. 
Specifically, we parametrize each branching node $t$ by a split function $s_{\theta_t}$ that routes an input to either its left or right child.

Hard context-to-node assignments would result in piece-wise constant functions with null gradients, making back-propagation ineffective.
Therefore, we make use of probabilistic relaxations, inspired by works such as that of~\citet{irsoy2012soft}, to make the tree learning problem differentiable, hence allowing the learning of the split parameters 
jointly by gradient descent so that they are optimal for retrieval.
In such a relaxation, a split function $s_{\theta_t}$ outputs a soft routing score rather than a hard left/right decision. This replaces hard routes through the tree with a probability distribution over the $|T|$ nodes of the tree, encoded by a vector $\mathbf{z}_i \in [0,1]^{|T|}$, where each entry $\mathbf{z}_{i,t}$ denotes the probability that input $i$ reaches node $t$ (see~\Cref{fig:architecture}).

\paragraph{Choice of Split Function.} 

A split function $s_{\theta_t}: \mathcal{X} \to [0, 1]$ can be implemented in several ways, as long as it outputs a routing score that determines the probability of routing an input left or right. 
Its simplest form is a linear projection, such as the one used in~\citet{pmlr-v139-zantedeschi21a}:
given a split node $t$, its left child $t_{\text{left}}$ and right child $t_{\text{right}}$, the split function is defined as the dot product $s_{\theta_t}(\xb_i) = \theta_t \xb^T$ (interpreting $\theta_t \in \mathcal{X}$ as a hyper-plane).
No matter the design of split function, the routing probabilities are finally computed as $\zb_{t_{\text{left}}} = \sigma(s_{\theta_t}(\xb_i))$ and $\zb_{t_{\text{right}}} = 1 - \sigma(s_{\theta_t}(\xb_i))$, with $\sigma(x) \vcentcolon= \frac{1}{1+ e^{-x}}$ the sigmoid function, to ensure a valid probability distribution.

While the linear split function is computationally efficient, it may lack the expressiveness to capture complex routing decisions. We therefore also explore neural network-based split functions. In particular, we propose a \textbf{cross-attention-based function} ~\citep{vaswani2017attention} that operates directly on the token-level representations from the encoder, where learnable tree embeddings attend to the input sequence $\xb_i \in \mathbb{R}^{n_d \times d}$ of $n_d$ token embeddings.
 The cross-attention split function (see Appendix~\ref{app:cross_attention_split_fn} for more details) is generally more expressive than a linear projection 
(as our ablation shows in \Cref{app:split}).
Indeed, the tree embeddings and projection matrices act as memories that store information from past query and context embeddings, useful for scoring the current inputs.

\paragraph{Tree Propagation}
Note that any split function defined above outputs a score that is independent of the scores from the other nodes.
However, the probability of reaching a node should be constrained by the probability of having reached its parent, the parent of its parent, and so on, to avoid degenerate trees where a descendant has higher probability of being reached than its ancestors.
The simplest way of enforcing such tree constraints is by propagating the scores of the ancestors down to the node of interest by multiplying the probabilities along the path.
We refer to this type of tree propagation as \textit{product propagation} and use it as default method in our experiments.
Given a node $t$ and its ancestors $\mathcal{A}_t$ (the  nodes along the path from the root to $t$), the probability of input $x_i$ reaching $t$ is $T(\xb_i)_t = \zb_{t} \prod_{a \in \mathcal{A}_t} \zb_{a}$.
Notice that the product propagation naturally enforces that the sum of node probabilities at any depth is always constant and equal to~1.
Alternatively, one can learn how to propagate probability scores through the tree.
We refer to this type of tree propagation as \textit{path consistency propagation} and describe it in ~\Cref{app:propagation}. 

To keep both training and inference efficient even for deep trees, we compute all split scores in parallel before applying the chosen propagation scheme.

\subsection{Training by Contrastive Learning}
The task now is to learn end-to-end the parameters $\Theta$ (that include those of the split and propagation functions, and of the tree embeddings), so that any query is routed optimally to the leaf that contains its corresponding ground-truth context. 
Such an optimal assignment is achieved when positive query-context pairs are independently routed similarly through the tree, leading to similar leaf assignments.
 However, optimizing solely for perfect positive assignments is likely to lead to the trivial solution where all contexts and queries are routed to a single leaf, resulting in a collapsed tree with a single active path.
To avoid such a solution, we optimize a contrastive objective that implicitly encourages the model to use the tree non-trivially by spreading unrelated queries and contexts across different leaves.
Building on contrastive learning literature~\citep{oord2018representation,radford2021learning}, we optimize the following Softmax-based InfoNCE loss,
\begin{equation}
    {-} \frac{1}{2|\cP|}\! \sum_{i=1}^{|\cP|}\! \left ( \!\log \frac{e^{\simi(\qb_i, \cb_i)}}{\sum_{j=1}^{|\cP|} e^{\simi(\qb_i, \cb_j)}} {+} \log \frac{e^{\simi(\cb_i, \qb_i)}}{\sum_{j=1}^{|\cP|} e^{\simi(\cb_i, \qb_j)}}\!\right ) \label{eq:constrative_loss}
\end{equation}
where $\cP$ is the set of query-context pairs, and we take the similarity $\simi: [0, 1]^{|\mathcal{T}_L|} \times [0, 1]^{|\mathcal{T}_L|} \to [-1,0]$ to be the negative Total Variation Distance (nTVD) between two leaf assignments: $\simi(\mathbf{a}, \mathbf{b}) = -\frac{1}{2} \sum_{l=1}^{|\mathcal{T}_L|} \left| a_l - b_l \right|$, where \( \mathbf{a} = (a_1, a_2, \ldots, a_{|\mathcal{T}_L|}) \) and \( \mathbf{b} = (b_1, b_2, \ldots, b_{|\mathcal{T}_L|}) \) are the leaf assignment distributions (i.e., the restriction of 
$T(x_i)$) to the leaf nodes, ignoring internal nodes), and \( |\mathcal{T}_L| \) is the number of leaf nodes in the tree.
In Eq.\eqref{eq:constrative_loss}, the left term encourages any query to have a leaf assignment similar to the assignment of its ground-truth context and different from any other context in the batch.
Conversely, the second term encourages contexts to be routed similarly to their positive queries and differently from their negative ones.
\vspace{-10pt}
\paragraph{Stochastic Depth Training} 
The contrastive loss in Eq.\eqref{eq:constrative_loss} is defined over leaf assignments. However, due to the hierarchical structure of the tree, optimizing at the leaf level implicitly shapes the assignments at intermediate levels as well, since each intermediate node's probability is the sum of its descendants' probabilities. To further strengthen these intermediate representations for inference (see \Cref{sec:inference_transparency_latency}), we introduce a training scheme where, at each iteration, a tree level is sampled uniformly at random and the contrastive loss is optimized with respect to assignments at that level. We refer to this as stochastic depth training, in contrast to constant depth training where optimization is performed exclusively at the leaf level. We denote models trained with the stochastic scheme as \textsc{ReTreever-Stochastic}.

\subsection{Inference, Transparency, and Latency}
\label{sec:inference_transparency_latency}
Once the tree is trained, we index new content by 
encoding each reference item with the encoder $E$ and then routing it through the tree $T$ to obtain node assignments.
\vspace{-10pt}
\paragraph{Coarse-to-Fine Representations}
Routing an encoded text through the tree yields hierarchical representations at multiple granularities. At each level, the assignment vector captures alignment with learned semantic groups, with dimensionality determined by the number of nodes at that level. The leaf-level representation has dimensionality $|\mathcal{T}_L|$, providing the finest semantic granularity. Intermediate levels offer progressively coarser representations (e.g., dimension 16 at level 4), enabling a flexible trade-off between representation cost and retrieval accuracy.

\vspace{-10pt}
\paragraph{Indexing and Search}
For retrieval, we consider two indexing strategies. In \textit{single-index search}
, we build a vector index based on the chosen level assignments, process the query, and return its nearest neighbors based on the nTVD metric used at training.
In \textit{multi-index search}, we exploit the implicit clustering induced by the tree: each document is assigned to the leaf node for which it has the highest probability, and we build a separate index for each leaf cluster. At query time, we route the query through the tree, select its top-$k$ leaves, and search only the corresponding indices. This reduces the number of documents examined, trading off recall for latency. We report data access percentage (fraction of the corpus searched) as a proxy for latency in our experiments.
\paragraph{Implicit Semantic Coherence}
Learning node assignment probabilities, as opposed to unconstrained features, 
naturally provides a hierarchical and soft clustering of the documents that can be inspected by the user.
We show in \Cref{sec:interpretability} that \ourmodel{}'s learned clusterings capture semantic groupings, where assigned documents share topics and keywords, even though \ourmodel{} does not directly optimize for semantic coherence.

\begin{table*}[h]
\centering
\caption{
Average retrieval latency (ms) and \textsc{nDCG@10} (\%) for tree-based methods on text retrieval tasks. \ourmodel{} strikes the best latency-performance trade-off among hierarchical retrievers.
}
\vspace{-5pt}
\label{tab:text-clustering}
\begin{center}
\begin{sc}
\resizebox{0.9\linewidth}{!}{%
\begin{tabular}{lcccccccc}
\toprule
 & \multicolumn{2}{c}{\textbf{NQ}} & \multicolumn{2}{c}{\textbf{HotpotQA}} & \multicolumn{2}{c}{\textbf{TopiOCQA}} & \multicolumn{2}{c}{\textbf{RepLiQA}} \\
\textbf{Method} 
& Lat. $\downarrow$ & nDCG${@}10$ $\uparrow$ 
& Lat. $\downarrow$ & nDCG${@}10$ $\uparrow$ 
& Lat. $\downarrow$ & nDCG${@}10$ $\uparrow$ 
& Lat. $\downarrow$ & nDCG${@}10$ $\uparrow$ \\
\midrule
Hier-Kmeans 
& 4.60 & 14.87
& 3.82 & 27.34
& 3.71 & 7.50
& 3.70 & 27.06 \\
\textsc{Hier-GMM}
& 5.42 & 11.15
& 5.06 & 17.19
& 3.81 & 4.54
& 8.67 & 14.03 \\
Raptor
& 1758.05 & 40.28
& 2246.51 & 59.37
& 701.61  & 18.69
& 1647.18 & 49.58 \\
\ourmodel{}
& 10.65 & \textbf{54.48}
& 10.01 & \textbf{85.04}
& 9.66  & \textbf{31.00}
& 9.47  & \textbf{80.23} \\
\bottomrule
\end{tabular}
}
\end{sc}
\end{center}
\end{table*}

\section{Retrieval Experiments}
\label{sec:experiments}
In this section, we assess the retrieval performance of \ourmodel{} at different representation levels and data access rates, as compared to flat and hierarchical retrieval methods.
Our results indicate that using the learned node assignments for retrieval (1) generally improves the representation power of the encoder at any representation level, (2) outperforms traditional approximate search even at low data access rates, (3) strikes the best latency-accuracy trade-off among hierarchical retrieval methods and (4) provides an inspectable and meaningful organization of content, making it much more transparent than similarity search based methods.

\subsection{Metrics and datasets}
To evaluate retrieval results, we measure the following standard metrics:
\textsc{Recall@k}, which assesses the proportion of ground-truth documents that are present within the top-k results returned by the retriever;
\emph{Normalized Discounted Cumulative Gain} at rank $k$ (\textsc{NDCG@k}) which accounts for ground-truth ranking, as relevant documents appearing earlier are considered more valuable.
We benchmark on four textual question-answering datasets (
 \textsc{Natural Questions} (\textsc{NQ})~\citep{kwiatkowski2019natural}, \textsc{HotpotQA}~\citep{yang2018hotpotqa}, \textsc{TopiOCQA}~\citep{adlakha2022topiocqa}, and \textsc{RepLiQA}~\citep{monteiro2024repliqa}), one image dataset (\textsc{ImageNet-1k}~\citep{deng2009imagenet}) and one speech dataset (\textsc{VoxCeleb2}~\citep{chung18b_interspeech}). 
For text retrieval datasets  composed of natural-language query–reference pairs, 
we follow the pre-processing procedure of \citet{monteiro2024xc}, except that for \textsc{HotpotQA} we concatenate all relevant contexts and discard irrelevant ones to obtain the reference.
For image and speech retrieval, we construct $m$ positive query-reference pairs per item, by coupling each item with another from the same class ($m=100$ for \textsc{ImageNet-1K} and $m=2$ for \textsc{VoxCeleb2}).
We make use of a validation set for model selection and hyperparameter tuning. 
For \textsc{RepLiQA}, we use the first four splits for training; from the final split, we randomly sample 10\% of the data for validation and use the remaining samples for testing.
For datasets with a validation set but no test partition, we use the validation set for testing and create a validation set by holding out 10\% of randomly selected training samples.
Then, for training \ourmodel{} and the baselines we make use of the training query-reference pairs, and for testing, we build the index with all the items from training, validation and test splits.

\subsection{Baselines}
\label{sec:baselines}
Across all considered input modalities (text, image, and audio), we evaluate \ourmodel{} against retrieval baselines based on standard encoders and coarse-to-fine representations. For text retrieval, we further include comparisons with hierarchical approaches.

\paragraph{\textsc{Encoder}.}  
For text retrieval, we compare against DistilBERT finetuned on MSMarco\footnote{\textit{sentence-transformers/msmarco-distilbert-cos-v5}}~\citep{reimers-2019-sentence-bert}; for image retrieval, the Visual Transformer Dinov2\footnote{\textit{facebook/dinov2-large}}~\citep{oquab2023dinov2}; for speech retrieval, the Audio Spectrogram Transformer (AST) model fine-tuned on AudioSet\footnote{\textit{MIT/ast-finetuned-audioset-10-10-0.4593}}~\citep{gong_psla,gong21b_interspeech}.
Additionally, we compare with dataset-specific versions of these encoders (\textbf{Encoder-MLP}), 
obtained by attaching a multilayer perceptron (MLP) projection head to each encoder and fine-tuning the resulting model separately on each dataset.
The MLP layer has input and output dimensions equal to the embedding size, and is trained using the contrastive loss in Eq.~\eqref{eq:constrative_loss}, with cosine similarity as the similarity measure.
We report a sensitivity analysis with respect to the choice of encoder in \cref{app:different_text_encoders} and \cref{app:compare_image_encoders}.
\paragraph{\textsc{Encoder-MRL}.}  We introduce a coarse-to-fine hierarchy into encoder embeddings by training the projection layer with the contrastive loss in a Matryoshka Representation Learning fashion~\citep{kusupati2022matryoshka} (see \cref{app:c2f_rep}). This training induces nested semantic granularity in the embedding space, making Encoder-MRL directly comparable to \ourmodel{}’s multi-level representations.  

\paragraph{Hierarchical clustering.} For text retrieval, we perform a top-down hierarchical k-means (\textbf{\textsc{Hier-Kmeans}}) or GMM (\textbf{\textsc{Hier-GMM}}) clustering, where documents are iteratively partitioned into two clusters at each level of the hierarchy based on the cosine similarity of their embeddings as extracted by an encoder. This recursive process continues until the hierarchy reaches the same depth of \ourmodel{}. A similar procedure is applied for building the index: contexts are greedily routed through the learned tree and assigned to the leaf they reach.
We apply tree search 
for both \textsc{Hier-Kmeans} and \textsc{Hier-GMM}, where instead of greedily traversing the tree (which would give poor performance), we compute a global search to find the leaf with the highest score: all split-node clustering models are evaluated, then their scores are propagated and aggregated top-down. Finally, the contexts assigned to the selected leaf are re-ranked based on the cosine similarity between query and context embeddings.

\noindent\textbf{\textsc{Raptor}}~\citep{sarthi2024raptor} is a hierarchical method that 
chunks long documents into shorter segments and then recursively builds a bottom-up tree via clustering and LLM-based summarization.
Although designed for document QA, \textsc{Raptor} builds a tree structure of contexts, making it a highly relevant baseline for our retrieval datasets, where each context can be viewed as one of those smaller segments.
For a fair assessment of its retrieval accuracy, we apply a similar tree search as for hierarchical clustering, by scoring nodes using the cosine similarity between query and summary embeddings. The contexts assigned to the selected leaf are then reranked again via the cosine similarity. Note that this is the only method we fit on the test corpus, as it cannot be applied to unseen documents and does not scale to our training sets.

\subsection{Implementation Details}

\begin{table*}  
\centering
\caption{Comparison of similarity search-based methods using fine representations.}
\label{tab:full-rep-main}

\small 

\begin{adjustbox}{max width=\textwidth}
\begin{tabular}{l@{\hskip 6pt}c@{\hskip 6pt}c@{\hskip 6pt}c@{\hskip 6pt}c@{\hskip 6pt}c@{\hskip 6pt}c}
\toprule
\textbf{Model} & \multicolumn{2}{c}{\textsc{NQ}} & \multicolumn{2}{c}{\textsc{ImageNet1K}} & \multicolumn{2}{c}{\textsc{VoxCeleb2}} \\
& \scriptsize{Rec${@}10$ $\uparrow$} & \scriptsize{nDCG${@}10$ $\uparrow$}
& \scriptsize{Rec${@}10$ $\uparrow$} & \scriptsize{nDCG${@}10$ $\uparrow$}
& \scriptsize{Rec${@}10$ $\uparrow$} & \scriptsize{nDCG${@}10$ $\uparrow$} \\
\midrule
Encoder & 57.53 & 42.15 & 92.89 & 73.60 & 85.65 & 44.94 \\
Encoder + Linear & 68.61 & 47.62 & 93.84 & 76.39 & 96.68 & 75.02 \\
Encoder + MLP & 75.10 & 52.23 & 93.49 & 79.66 & 96.25 & 73.38 \\
Encoder + MRL & 67.10 & 46.60 & 93.87 & 76.29 & 95.61 & 70.27 \\
\midrule
Retreever - Stochastic & 77.36 & 54.48 & 94.34 & 78.82 & 99.31 & 92.60 \\
Retreever              & \textbf{77.50} & \textbf{54.58} & \textbf{94.41} & \textbf{79.68} & \textbf{99.43} & \textbf{93.48} \\
\bottomrule
\end{tabular}%
\end{adjustbox}
\end{table*}

For all methods, we use representation size of $1024$ and, for text retrieval, $512$-token input truncation. 
\ourmodel{} has depth $10$ and uses $8$ attention heads for the cross-attention split. 
When not specified otherwise, we use exact search with cosine similarity (\textsc{Encoder}) or nTVD (\ourmodel{}) for top-$k$ retrieval. For coarse-to-fine comparison, we use node assignments at level $h$ for \ourmodel{}, and the first $2^h$ dimensions of Encoder embeddings to compare at the same representation size. See \cref{app:c2f_rep} for exhaustive details.

\subsection{Experimental Results}

\paragraph{Comparison with tree based methods}\Cref{tab:text-clustering} compares tree-based methods' retrieval performance and inference time, across all text datasets. 
Remarkably, \ourmodel{} achieves the best retrieval accuracy while having a runtime several order faster than Raptor's, and similar to the other hierarchical retrieval methods. 
On average, \ourmodel{} processes each query in just $~10$ ms.
Compared to \textsc{Raptor}, it is about $160\times$ faster 
while obtaining performance boosts of around $30\%$, despite not making LLM calls for building the tree.
Finally, \ourmodel is about $2\times$ slower than \textsc{Hier-Kmeans} and \textsc{Hier-GMM} but achieves at least $2.5\times$ better performance. 
The increased latency is due to our model having higher capacity split functions, as compared to dot products for computing cosine similarity. 
Overall, these results provide strong evidence for learning tree structures optimal for retrieval.

\paragraph{Fine Representation Comparison} 
Table~\ref{tab:full-rep-main} reports retrieval performance across modalities for all representation-based methods. We observe similar results on the other text datasets in \cref{tab:full-rep-app}.
First, \ourmodel{} consistently achieves the best results across both recall and nDCG, substantially outperforming \textsc{Encoder}. This demonstrates that learning a tree does not compromise the encoder's representational power, and that transparency does not come at the cost of retrieval performance.
Second, comparing \textsc{Encoder} with other baselines shows that finetuning the encoder on task-specific data unsurprisingly improves performance. The superior results of our finetuning strategy can be attributed to its use of higher-capacity split functions (attention-based versus linear or MLP adaptors).
Finally, we observe that explicitly training for high-quality coarse representations (\textsc{Encoder-MRL} and \textsc{Retreever-Stochastic}) slightly degrades fine-level representation power. We further investigate the impact of coarse-to-fine optimization strategies in what follows.

\begin{figure}[h]
    \centering
    \vspace{2mm}
    \includegraphics[width=0.75\textwidth]{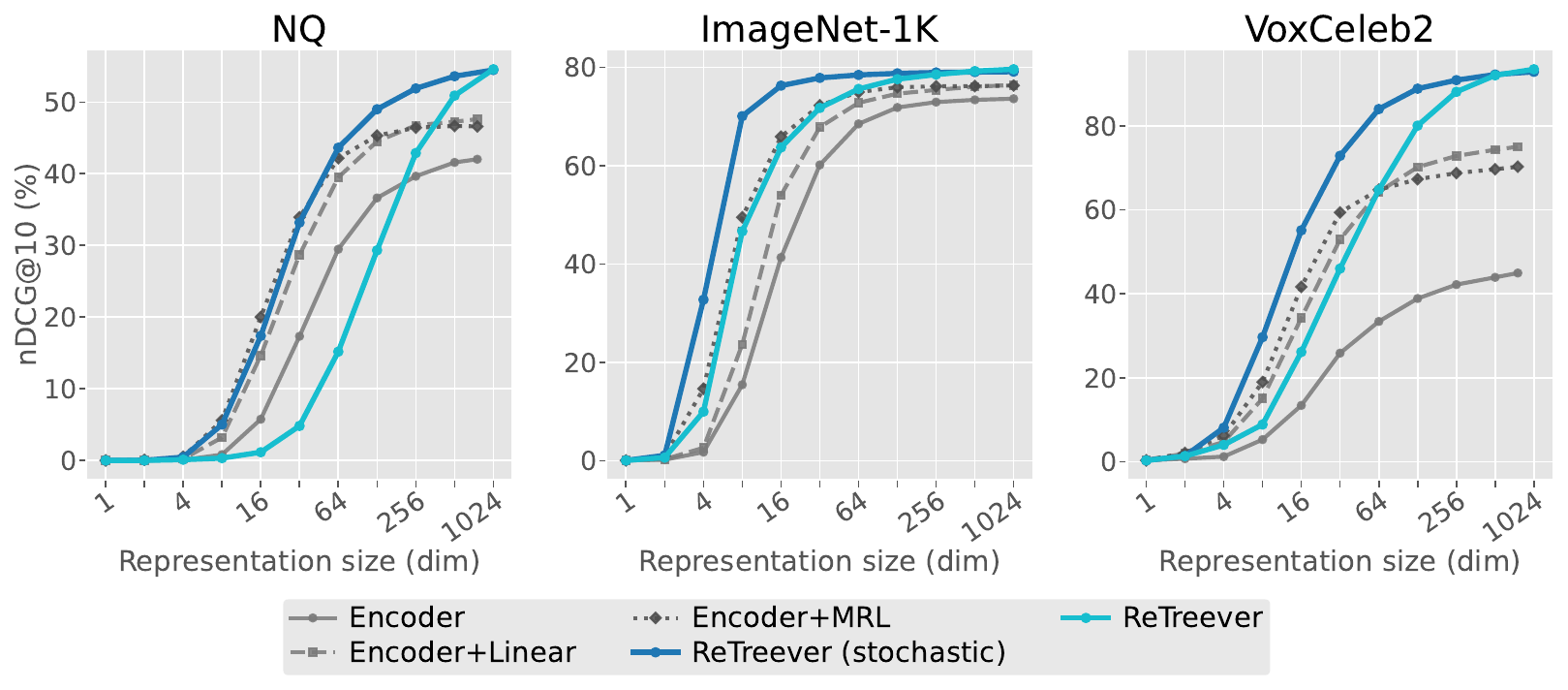}
    \caption{\textsc{nDCG@10} (\%) vs.\ representation size across modalities, using a single index. For each dataset, we use an encoder appropriate to its data modality, as described in Section~\ref{sec:baselines}.}
    \label{fig:sparse-rep-main}
\end{figure}

\paragraph{Coarse-to-fine Representation} 
In \cref{fig:sparse-rep-main}, we compare retrieval performance across text, image and speech datasets, depending on the representation size.
We observe that \textsc{\ourmodel{}-Stochastic} delivers the strongest coarse representations, as measured by \textsc{nDCG@10}, consistently across modalities, improving over all baselines at any representation size. In particular, it outperforms \textsc{Encoder-MRL}, the other method explicitly optimizing for high quality coarse representations.
Strong coarse representations cut inference cost massively by 75\% while retaining 95\%  of \textsc{nDCG@10} (averaged across all datasets) (see \cref{app:rep_size_vs_latency}, \cref{fig:latency_tradeoff}).
Finally, imposing coarse-to-fine constraints degrades performance of the finer representations across all datasets, as seen when moving from \textsc{Encoder-Linear} to \textsc{Encoder-MRL}, and slightly from \textsc{ReTreever} to \textsc{ReTreever-Stochastic} (see \cref{tab:full-rep-main}).

\begin{figure}
    \centering
    \includegraphics[width=0.75\textwidth]{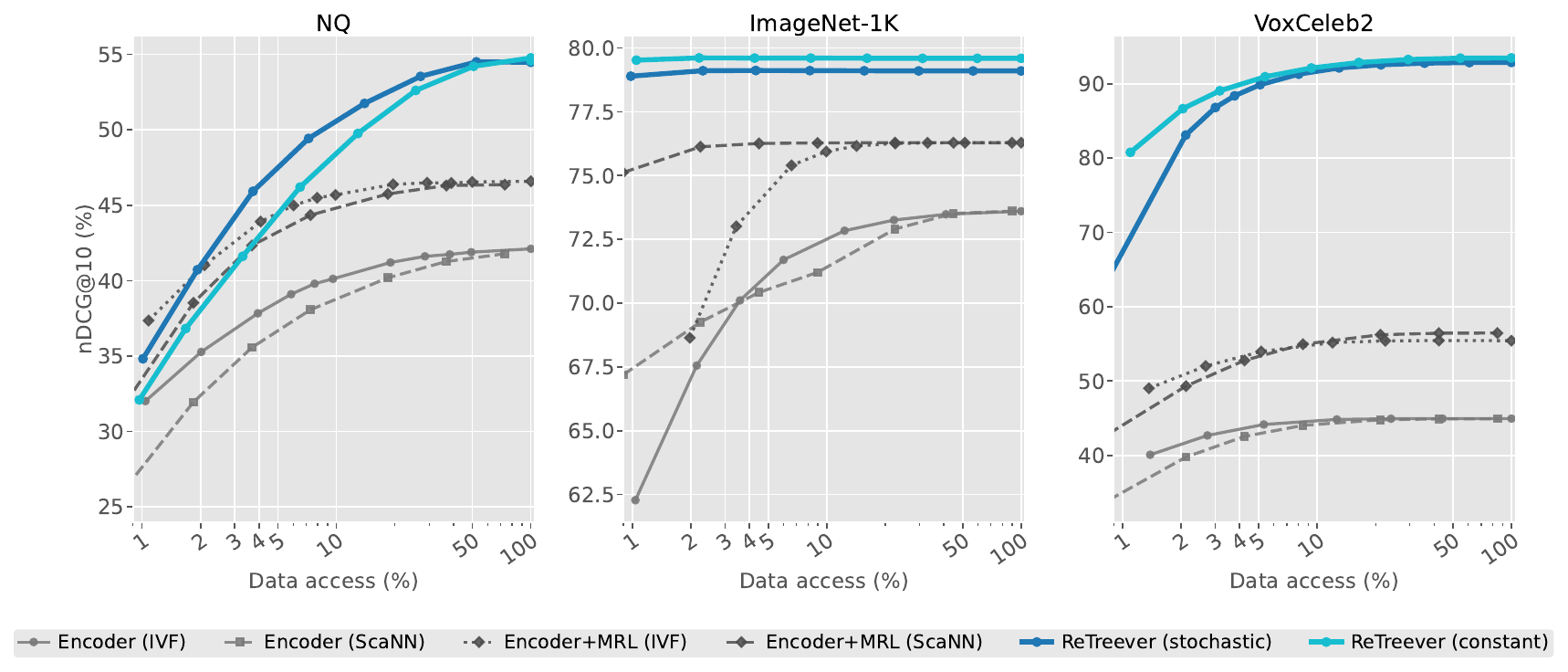}
    \caption{\textsc{nDCG@10} (\%) vs (\%) Data Accessed across modalities. \ourmodel{} is evaluated using leaf-level indices, while baselines make use of the approximate searches: IVF and SCANN. For each dataset, we use an encoder appropriate to its data modality, as described in Section~\ref{sec:baselines}.}
    \label{fig:multi_index_results}
\end{figure}

\paragraph{Approximate Data Search} 
In \cref{fig:multi_index_results}, we study retrieval performance as we progressively restrict the amount of data that can be accessed during similarity search.
On all datasets, we observe significant performance degradation for all methods with data budgets below $30\%$, with the exception of \textsc{ImageNet-1K} where \ourmodel{} and \textsc{Encoder+MRL} (ScaNN) prove to be robust to extremely low data regimes.
In general, our tree-based approach consistently outperforms \textsc{Encoder} and \textsc{Encoder-MRL} combined with standard approximate search methods across all budgets. 
Remarkably, \ourmodel{} performs well even though it is not designed as an approximate search method. Unlike IVF and ScaNN, which use task-agnostic clustering to create partitions, our method learns hierarchical partitions directly optimized for retrieval. This is further evidence that our method learns a meaningful organization of the data.
Finally, we observe that \textsc{ReTreever} and \textsc{ReTreever-Stochastic} perform similarly over all data access budgets, except on \textsc{VoxCeleb2} where the stochastic variant is substantially stronger at low data-access regimes.  
Overall, these results demonstrate that leveraging the tree's partitions achieves at least $2\times$ latency reduction with minimal impact on retrieval quality.

\section{Inspecting the Tree}
\label{sec:interpretability}

\begin{figure*}[t]
\centering

\begin{minipage}[b]{0.38\textwidth}
  \centering
  \includegraphics[width=\linewidth]{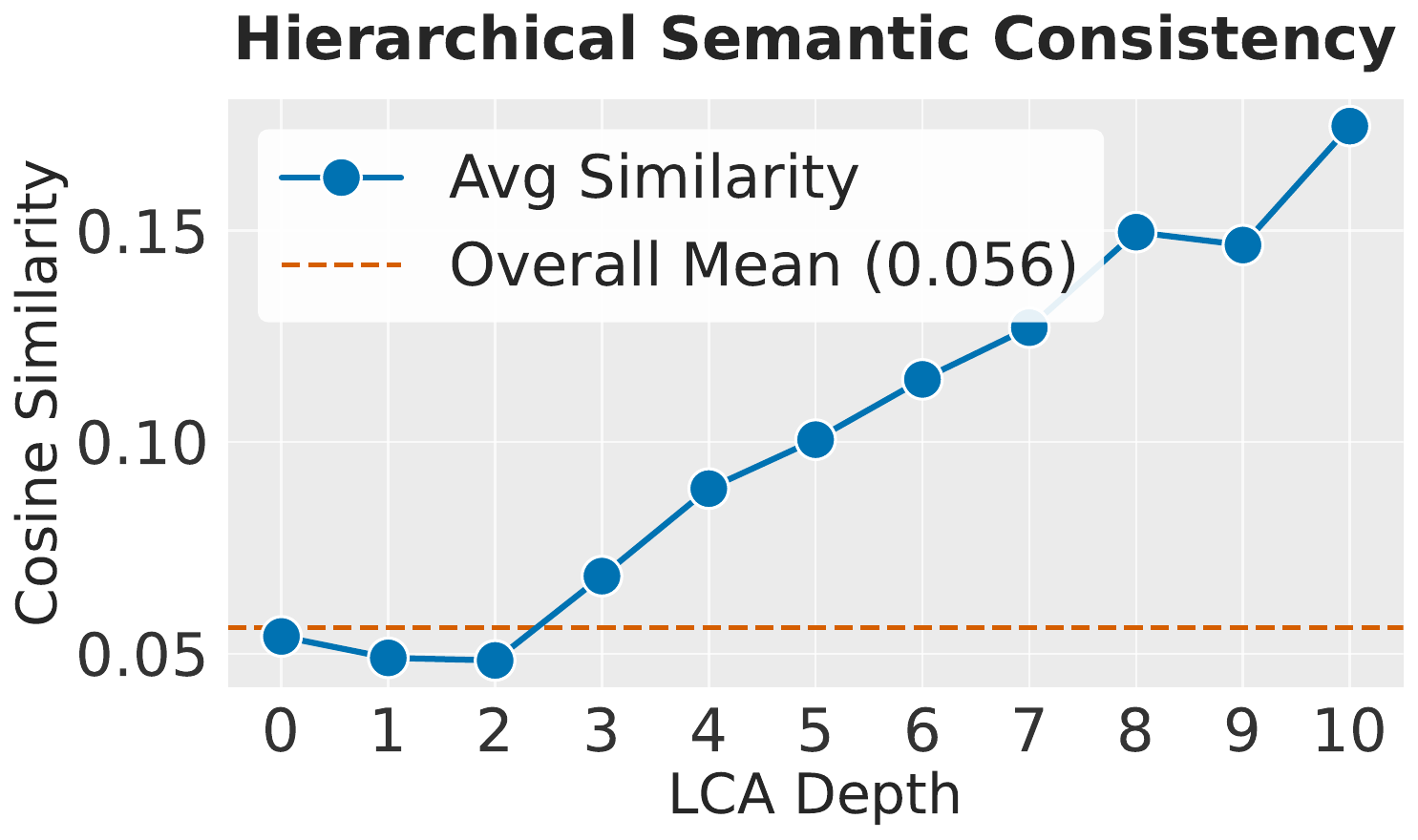}
  \textbf{\small (a)}\;{\small Semantic consistency}
\end{minipage}
\begin{minipage}[b]{0.4\textwidth}
  \centering
  \includegraphics[width=\linewidth]{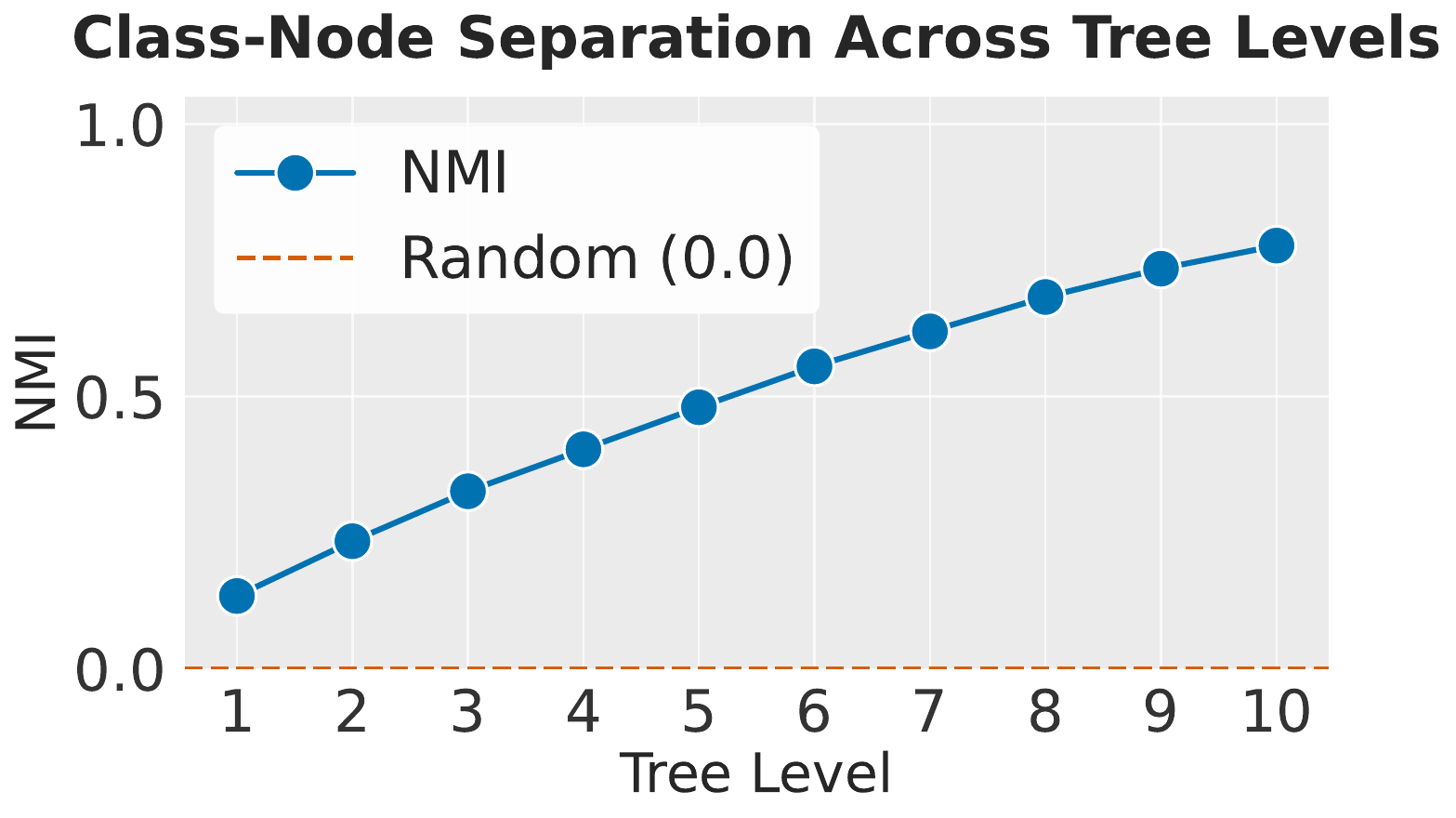}
  \textbf{\small (b)}\;{\small Node purity}
\end{minipage}
\caption{Learned trees exhibit semantic structure. (a) Documents with deeper common ancestors are more semantically similar (NQ). (b) Node purity improves with tree depth (\textsc{ImageNet-1K}).}
\label{fig:hierarchy_analysis}
\end{figure*}

\begin{figure}[H]
\centering
\begin{tabular}{cc}
\includegraphics[width=0.40\textwidth, angle=-90, origin=c]{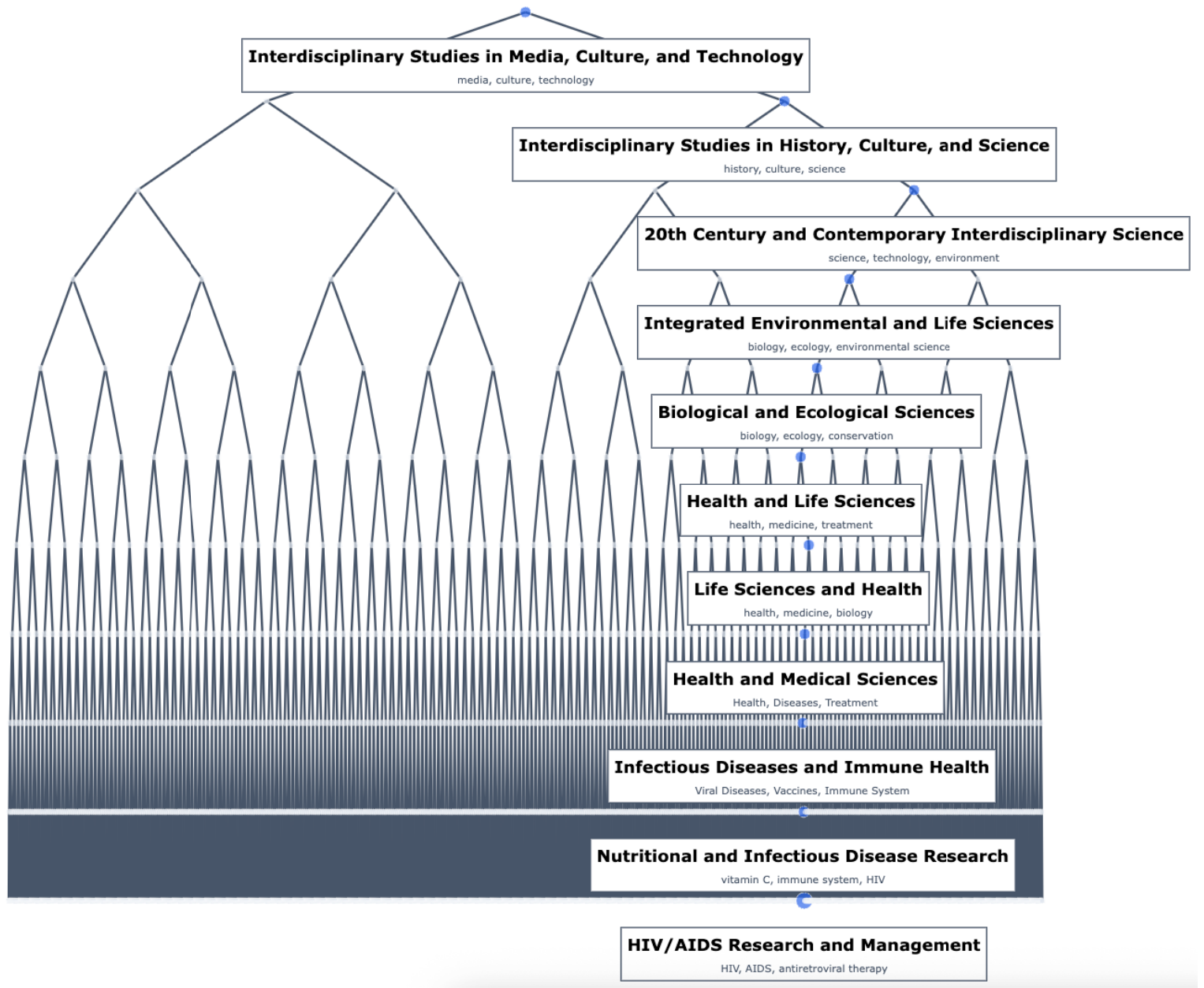} &
\includegraphics[width=0.40\textwidth, angle=-90, origin=c]{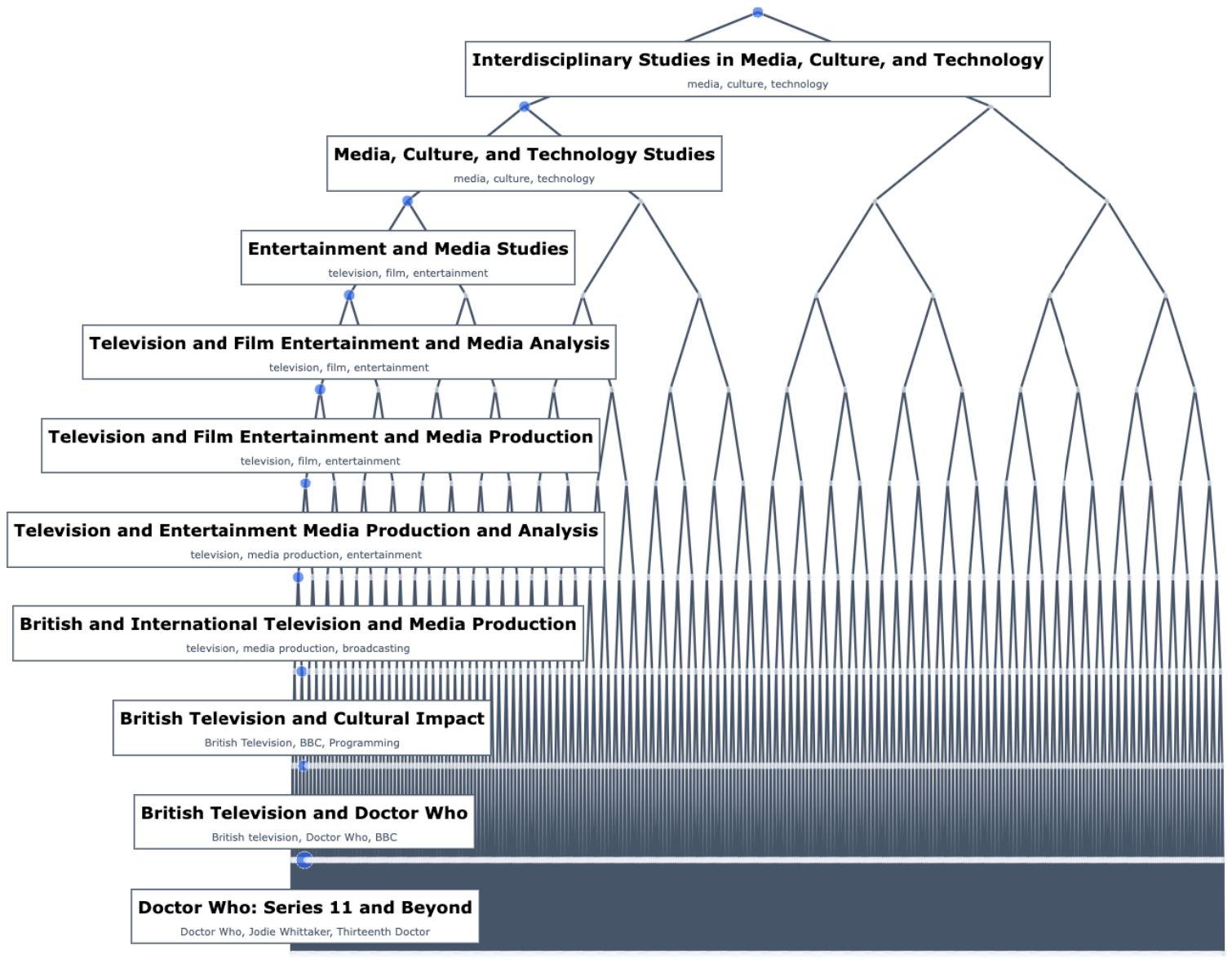} 
\end{tabular}
\caption{Examples of hierarchical document organization. \textit{Left}: Science $\rightarrow$ Biology $\rightarrow$ Health $\rightarrow$ Medicine $\rightarrow$ Infectious Diseases $\rightarrow$ HIV. \textit{Right}: Media $\rightarrow$ Entertainment $\rightarrow$ Television $\rightarrow$ British Television $\rightarrow$ Doctor Who $\rightarrow$ Doctor Who (Season 11).}
\label{fig:hierarchichal_org_examples_main_paper}
\end{figure}

By learning node assignment probabilities, \ourmodel{} has the appealing side-effect of providing an organization of reference documents, which serves as an interface for inspecting the model and analyzing the contexts. 
In this section, we probe \ourmodel{}'s organization via several tools and report evidence that semantic groupings naturally emerge when optimizing for retrieval. 
\paragraph{Visual Inspection Tool} We release an \href{https://euphonious-brigadeiros-cf5172.netlify.app/}{interactive visualization website} that renders trees learned by \ourmodel{} on text and image datasets for inspection. Examples like those in \cref{fig:hierarchichal_org_examples_main_paper} illustrate how \ourmodel{} refines concepts with depth: one branch in the tree narrows from Science $\rightarrow$ Biology $\rightarrow$ Health $\rightarrow$ Medicine to Infectious Diseases and ultimately HIV, while another specializes from Media $\rightarrow$ Entertainment $\rightarrow$ Television to British Television and finally to \textit{Doctor Who} (Season 11). These paths show that the tree captures both broad topical structure and highly specific subtopics at deeper levels. See \Cref{app:viz_tool}, for details and further examples (\Cref{fig:viz_tool_features}-\Cref{fig:elephants}).
\paragraph{Tree Congruence}
We first examine whether \ourmodel{} learns representations that are congruent with its tree structure,
despite being trained solely on query-context alignment labels. 
In \cref{fig:hierarchy_analysis}, we provide quantitative evidence that \ourmodel{} learns trees that exhibit semantic structure across modalities. 
For NQ, we study whether documents assigned to nearby nodes are more similar, on average, than documents assigned to different subtrees.
We compute the cosine similarity between document embeddings (extracted with DistilBERT) grouped by the depth 
of their \emph{lowest common ancestor} (LCA), the deepest node in the tree that is an ancestor of both contexts.
The hypothesis is that contexts routed to different nodes at earlier layers (i.e. with a shallow LCA closer to the root) 
should be thematically more distinct than those separated deeper in the tree. 
The strong positive correlation shows that documents sharing deeper (i.e. closer to the leaves) common ancestors are semantically more similar. 
For \textsc{ImageNet-1K}, we study whether images from the same class are routed similarly through the tree.
Given the tree paths of the images, we compute the \emph{Normalized Mutual Information} (NMI) between their class distribution and their tree level assignments. 
NMI increases with level depth, indicating that class purity increases with depth and demonstrating progressively finer semantic separation. 
See Appendix~\ref{app:quant_metrics} for an extended analysis. 
\paragraph{Topics and Keywords Coherence}
To further inspect the learned structure in a human-understandable way, we resort to topic models for extracting and analyzing the distribution of topics and keywords of the contexts assigned to nodes across the tree. 
We present
a visualization of the \ourmodel{} tree trained on \textsc{NQ}, highlighting the topics and keywords associated with several nodes (Figure~\ref{fig:transparency}).
For each node $t$, we collect all contexts assigned to the leaves of the subtree rooted at $t$ and extract topics and keywords using the method from \citet{kilgarriff2009simple}.
A quick inspection of these keywords reveals that the learned hierarchy reflects the semantic relationships among contexts. 
For example, the node whose contexts are on the topic of \emph{media}  (node 5) has child nodes focused on \emph{publishing} and \emph{TV}. Furthermore, the path from node 5 to node 54 illustrates a consistent refinement of topics and keywords, progressing from \emph{Media} down to \emph{Television seasons}.
 
\section{Discussion}
This work presents \ourmodel{}, a hierarchical retriever system that generates tree-based representations optimized for retrieval. By aligning coarse-to-fine representations with a learned hierarchy, \ourmodel{} enables flexible trade-offs between representation size, inference latency, and retrieval accuracy, while remaining competitive with flat retrieval methods and offering substantially improved efficiency at low data-access regimes. Beyond efficiency, the learned tree provides an inspectable organization of the corpus, allowing practitioners to examine intermediate groupings, and offering some degree of transparency that is largely absent from standard embedding-based retrieval systems.
Several promising directions emerge from this work. While \ourmodel{} currently operates on top of a frozen encoder and a single learned hierarchy per dataset, extending the framework to support adaptive or incremental tree updates could improve its applicability to dynamic or evolving corpora. Learning variable-depth or variable-width trees may further improve efficiency by allocating capacity where it is most needed, while tighter integration between encoder and tree learning could strengthen semantic alignment.

\bibliography{references}
\bibliographystyle{servicenow}

\appendix
\onecolumn

\section{\ourmodel{}'s Additional Design Options}
\ourmodel{} can be instantiated in many different ways. In this section we report a selection of the design choices we explored in our experiments and that are implemented in our codebase.

\subsection{Split Functions}
\label{app:split}
A split function $s_{\theta_t}: \mathcal{X} \to [0, 1]$ determines the routing probability of an input $\xb_i \in \mathcal{X}$ through node $t$. The primary requirement for a split function is that it outputs a scalar $\in \mathbb{R}$, based on which we compute the probabilities  of routing the input to node $t$'s left or right children $t_{\text{left}}$ and $t_{\text{right}}$ (see \Cref{app:propagation}).
Below, we describe different types of split functions that can be used.

\subsubsection{Linear Split Function}
The simplest form of a split function is a linear projection, similar to the approach used in~\cite{pmlr-v139-zantedeschi21a}. For a given split node $t$, with left and right children $t_{\text{left}}$ and $t_{\text{right}}$, the split function is defined as:
\begin{equation}
    s_{\theta_t}(\xb_i) = \theta_t^\top \xb_i,
\end{equation}
where $\theta_t \in \mathcal{X}$ represents a learnable hyperplane.
Each split node learns a separate hyperplane, and there are no shared parameters across the tree.

\subsubsection{MLP Split Function}
A more expressive alternative is to use a learnable neural network, modeled as a Multi-Layer Perceptron (MLP). This MLP $S_{\Theta}: \mathcal{X} \to \mathbb{R}^{|\mathcal{T}_B|}$ maps an input $\xb_i$ to a routing probability for each branching node in the tree, where $\mathcal{T}_B$ is the set of branching nodes (i.e. non leaf nodes). The MLP consists of multiple layers with nonlinear activations, such as ReLU, and incorporates dropout for regularization. Unlike the linear split function, which maintains separate parameters per node, the MLP split function shares parameters across different nodes while still allowing for node-specific learning.

\subsubsection{Cross-Attention Split Function}
\label{app:cross_attention_split_fn}
While the linear and MLP split functions operate on dense passage embeddings for the entire document, the cross-attention split function allows us to leverage token-level embeddings for more expressive routing. This method utilizes a cross-attention mechanism between learnable node embeddings and text tokens to determine the routing probabilities at each split node.

Let the input text $\xb_i \in \mathbb{R}^{n_d \times d_{\text{emb}}}$ consist of $n_d$ embedded tokens, encoded by the encoder $E$. Instead of a simple projection, we introduce learnable level embeddings $\eb_l \in \mathbb{R}^{n_e \times d_{\text{emb}}^{\prime}}$ for each tree level $l$. These level embeddings interact with the text tokens via a cross-attention mechanism, where the level embeddings serve as queries, and the text tokens provide keys and values. 

We define the attention mechanism as follows:
\begin{equation}
    \text{Attention}(\mathbf{Q}, \mathbf{K}, \mathbf{V}) = \text{softmax}\left(\frac{\mathbf{Q} \mathbf{K}^\top}{\sqrt{d_k}}\right) \mathbf{V},
\end{equation}
where:
\begin{align}
    \mathbf{Q} &= \eb_l \mathbf{W}_q^\top, \\
    \mathbf{K} &= \xb_i \mathbf{W}_k^\top, \\
    \mathbf{V} &= \xb_i \mathbf{W}_v^\top.
\end{align}
Here, $\mathbf{W}_q \in \mathbb{R}^{d_k \times d_{\text{emb}}^{\prime}}$, $\mathbf{W}_k \in \mathbb{R}^{d_k \times d_{\text{emb}}}$, and $\mathbf{W}_v \in \mathbb{R}^{d_k \times d_{\text{emb}}}$ are learnable projection matrices shared across the tree, while $d_k$ represents the dimension of the projected queries and keys.
This formulation describes a single-head attention mechanism but can be naturally extended to multi-head attention by introducing independent projection matrices for each head and concatenating the resulting outputs.

\begin{figure}[t]
    \centering
    \includegraphics[width=0.90\textwidth]{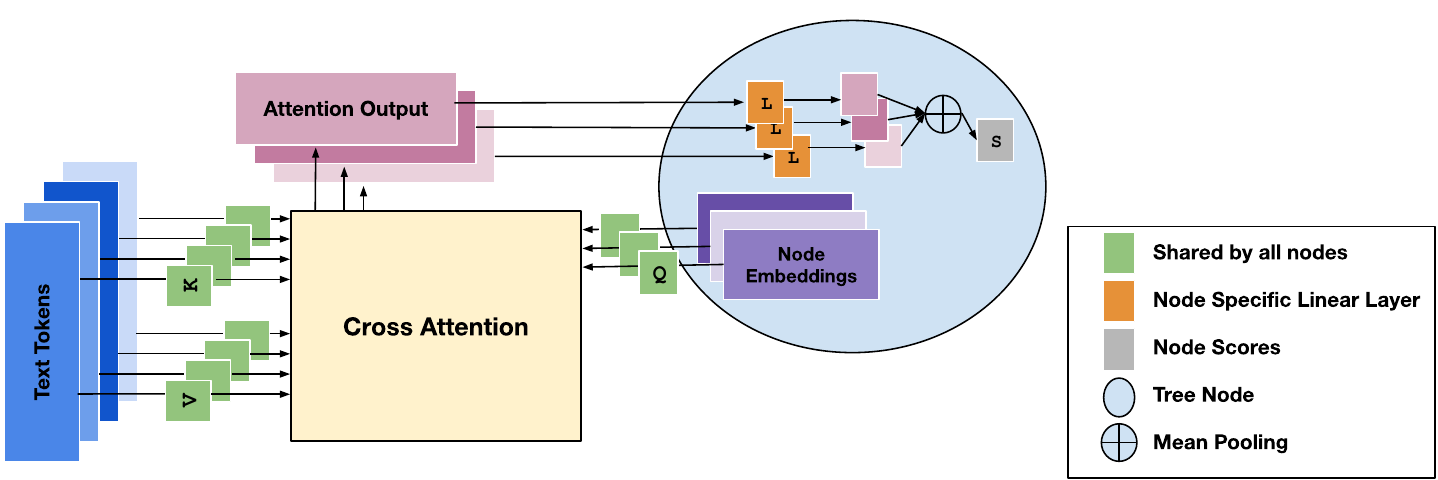}
    \caption{\ourmodel{}'s cross-attention split function with node scoring done by a per node linear map followed by a mean of scores.}
    \label{fig:cross_attn_split_fn_with_scoring}
\end{figure}

The transformed (contextualized) level embeddings are then converted into node scores using a node-specific linear function. When multiple level embeddings are available at a given level, we apply the same node-specific linear function to each embedding, and obtain the final score for the node by averaging the resulting scores. This aggregation is shown in \autoref{fig:cross_attn_split_fn_with_scoring}. This mechanism significantly increases the expressivity of the split function compared to a simple linear projection. The node embeddings and projection matrices act as memory representations, storing information from past query and context embeddings, which enhances the model's ability to score inputs effectively.

We compare the effect of different split functions on retrieval performance across various datasets in \cref{tab:split-function-ablation}. Notably, the cross-attention split function achieves the best retrieval metrics across all datasets, while the MLP split function worse than the linear one.
This surprising result can be explained by highlighting that both cross-attention and linear splits learn virtual embeddings (the level embeddings combined with node projection functions in the cross-attention and the hyper-plane in the linear one) and compare them with the input. Further investigation would be required to confirm this speculation.

\begin{table}[h]
\vspace{10pt}
\centering
\caption{Effect of split-function choices on full-representation retrieval \textsc{Recall@10} (\%) and \textsc{nDCG@10} (\%) across three datasets, with best results per metric and dataset in bold.}
\label{tab:split-function-ablation}
\begin{center}
\begin{sc}
\begin{tabular}{l@{\hskip 6pt}c@{\hskip 6pt}c@{\hskip 6pt}c@{\hskip 6pt}c@{\hskip 6pt}c@{\hskip 6pt}c}
\toprule
\textbf{Split function} & \multicolumn{2}{c}{\textsc{NQ}} & \multicolumn{2}{c}{\textsc{ImageNet1K}} & \multicolumn{2}{c}{\textsc{VoxCeleb2}} \\
& \scriptsize{Rec${@}10$ $\uparrow$} & \scriptsize{nDCG${@}10$ $\uparrow$}
& \scriptsize{Rec${@}10$ $\uparrow$} & \scriptsize{nDCG${@}10$ $\uparrow$}
& \scriptsize{Rec${@}10$ $\uparrow$} & \scriptsize{nDCG${@}10$ $\uparrow$} \\
\midrule
Linear            & 69.71 & 48.60 & 93.58 & 78.43 & 96.17 & 71.58 \\
MLP               & 57.69 & 39.06 & 92.85 & 75.60 & 82.17 & 39.21 \\
Cross-attention   & \textbf{77.50} & \textbf{54.58} & \textbf{93.92} & \textbf{79.59} & \textbf{99.43} & \textbf{93.48} \\
\bottomrule
\end{tabular}
\end{sc}
\end{center}
\end{table}

\subsection{Discussion on Cross-Attention Split Function}
\paragraph{Use of level embeddings as queries} The cross-attention split function establishes a structural parallel to the encoder-decoder framework found in standard Transformer architectures. In this analogy, the text encoder provides the keys and values, while the learnable level embeddings $e_l$ function analogously to decoder queries. Rather than generating text, these embeddings act as learnable semantic probes that query the document to extract features relevant to hierarchical routing. We believe that by using the same level embeddings across the entire corpus during training, the model effectively learns to store a ``global memory'' of the dataset's distribution within the tree parameters. We hypothesize that this allows the routing mechanism to perform a content-dependent summarization of the document, thereby asking which token-level features correspond to the global concepts associated with that specific depth, before specializing these summaries into node-specific scores through a final linear projection. This separation of concerns allows the model to maintain a shared semantic context at each level, while retaining the granularity required for individual routing decisions. Further investigation is required to confirm our hypothesis.

\subsection{Number of Heads in The Cross-Attention Mechanism}

We show in \cref{app:cross_attention_split_fn} that the cross-attention split function naturally extends to a multi-head setting. Here, we evaluate different numbers of attention heads across datasets. As shown in \cref{tab:num-heads-ablation}, $16$ heads achieve the best performance on $3$ of the $6$ dataset–metric combinations. We therefore use $16$ heads in all experiments.

\begin{table}[H]
\vspace{1em} 
\centering
\caption{Effect of the number of cross-attention heads on full-representation retrieval \textsc{Recall@10} (\%) and \textsc{nDCG@10} (\%) across three datasets, with best results per metric and dataset in bold.}
\label{tab:num-heads-ablation}
\begin{center}
\begin{sc}
\begin{tabular}{l@{\hskip 6pt}c@{\hskip 6pt}c@{\hskip 6pt}c@{\hskip 6pt}c@{\hskip 6pt}c@{\hskip 6pt}c}
\toprule
\textbf{\# Heads} & \multicolumn{2}{c}{\textsc{NQ}} & \multicolumn{2}{c}{\textsc{ImageNet1K}} & \multicolumn{2}{c}{\textsc{VoxCeleb2}} \\
& \scriptsize{Rec${@}10$ $\uparrow$} & \scriptsize{nDCG${@}10$ $\uparrow$}
& \scriptsize{Rec${@}10$ $\uparrow$} & \scriptsize{nDCG${@}10$ $\uparrow$}
& \scriptsize{Rec${@}10$ $\uparrow$} & \scriptsize{nDCG${@}10$ $\uparrow$} \\
\midrule
2  & 73.14 & 50.09 & 93.85 & 79.02 & 99.26 & 92.23 \\
4  & 73.73 & 50.93 & \textbf{94.02} & 79.98 & 99.29 & 92.25 \\
8  & 74.43 & 51.22 & 93.93 & 80.51 & 99.37 & 92.85 \\
16 & \textbf{77.50} & \textbf{54.58} & 93.92 & 79.59 & \textbf{99.43} & \textbf{93.48} \\
32 & 75.01 & 51.69 & 93.82 & \textbf{80.59} & 99.27 & 92.07 \\
\bottomrule
\end{tabular}
\end{sc}
\end{center}
\end{table}

\subsection{Choice of Tree Depth}
\label{app:tree_depth_choice}
We also study the effect of tree depth on retrieval performance in \cref{tab:tree-depth-ablation}. Increasing depth generally improves performance, and the largest depth we evaluate ($12$) achieves the best results for most dataset–metric combinations. However, on \textsc{VoxCeleb2} we observe signs of overfitting at depth $12$, where shallower trees (depth $8$ or $10$) perform better. Based on this tradeoff, and to keep the finest-level representation size comparable to standard encoders, we use depth $10$ in our experiments, yielding leaf-level representations of length $1024$.

\begin{table}[h]
\vspace{1em} 
\centering
\caption{Effect of tree depth on full-representation retrieval \textsc{Recall@10} (\%) and \textsc{nDCG@10} (\%) across three datasets, with best results per metric and dataset in bold.}
\label{tab:tree-depth-ablation}
\begin{center}
\begin{sc}
\begin{tabular}{l@{\hskip 6pt}c@{\hskip 6pt}c@{\hskip 6pt}c@{\hskip 6pt}c@{\hskip 6pt}c@{\hskip 6pt}c}
\toprule
\textbf{Depth} & \multicolumn{2}{c}{\textsc{NQ}} & \multicolumn{2}{c}{\textsc{ImageNet1K}} & \multicolumn{2}{c}{\textsc{VoxCeleb2}} \\
& \scriptsize{Rec${@}10$ $\uparrow$} & \scriptsize{nDCG${@}10$ $\uparrow$}
& \scriptsize{Rec${@}10$ $\uparrow$} & \scriptsize{nDCG${@}10$ $\uparrow$}
& \scriptsize{Rec${@}10$ $\uparrow$} & \scriptsize{nDCG${@}10$ $\uparrow$} \\
\midrule
6  & 64.51 & 42.98 & 94.06 & 77.84 & 98.54 & 86.14 \\
8  & 72.72 & 50.19 & \textbf{94.08} & 79.63 & 99.27 & \textbf{92.36} \\
10 & 73.81 & 50.85 & 94.01 & 80.62 & \textbf{99.31} & 92.27 \\
12 & \textbf{75.55} & \textbf{52.30} & 93.69 & \textbf{81.14} & 99.07 & 89.60 \\
\bottomrule
\end{tabular}
\end{sc}
\end{center}
\end{table}

\subsection{Number of Level Embeddings}
\label{app:abaltion_level_emb_num}
We also ablate the number of learnable level embeddings per tree level. As described in \cref{app:cross_attention_split_fn}, our cross-attention split function maintains $n_\ell$ learnable embeddings at each level; for a depth-$10$ tree this corresponds to $10 n_\ell$ level embeddings in total. These embeddings are contextualized via cross-attention with the input, and then mapped to node scores through node-specific projection functions (averaging scores when multiple level embeddings are used). \Cref{tab:embeddings-per-level-ablation} shows that increasing $n_\ell$ generally improves retrieval performance across datasets, but with diminishing returns beyond a small number of embeddings. Based on this tradeoff, we use $n_\ell=8$ embeddings per level in all experiments.

\begin{table}[H]
\vspace{1em} 
\centering
\caption{Effect of the number of embeddings per level on full-representation retrieval \textsc{Recall@10} (\%) and \textsc{nDCG@10} (\%) across three datasets, with best results per metric and dataset in bold.}
\label{tab:embeddings-per-level-ablation}
\begin{center}
\begin{sc}
\begin{tabular}{l@{\hskip 6pt}c@{\hskip 6pt}c@{\hskip 6pt}c@{\hskip 6pt}c@{\hskip 6pt}c@{\hskip 6pt}c}
\toprule
\textbf{\# Embeddings / level} & \multicolumn{2}{c}{\textsc{NQ}} & \multicolumn{2}{c}{\textsc{ImageNet1K}} & \multicolumn{2}{c}{\textsc{VoxCeleb2}} \\
& \scriptsize{Rec${@}10$ $\uparrow$} & \scriptsize{nDCG${@}10$ $\uparrow$}
& \scriptsize{Rec${@}10$ $\uparrow$} & \scriptsize{nDCG${@}10$ $\uparrow$}
& \scriptsize{Rec${@}10$ $\uparrow$} & \scriptsize{nDCG${@}10$ $\uparrow$} \\
\midrule
1 & 73.71 & 50.73 & 93.95 & 78.23 & 98.31 & 92.29 \\
2 & 73.69 & 50.56 & 93.91 & 78.19 & 98.98 & 92.05 \\
4 & \textbf{74.67} & 51.39 & \textbf{94.03} & 79.28 & 99.27 & 92.66 \\
8 & 74.49 & \textbf{51.46} & 93.87 & \textbf{80.95} & \textbf{99.33} & \textbf{92.79} \\
\bottomrule
\end{tabular}
\end{sc}
\end{center}
\end{table}

\section{Coarse-to-Fine Representations Across Datasets}
\label{app:c2f_rep}
\begin{figure}[H]
\centering
\begin{tabular}{ccc}
\includegraphics[width=0.32\textwidth]{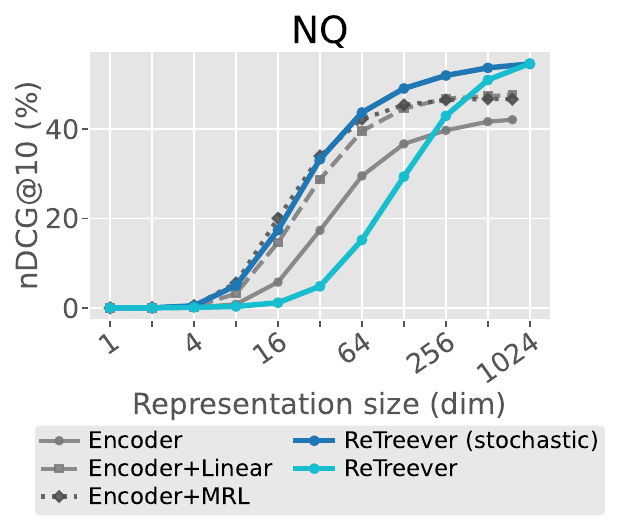} &
\includegraphics[width=0.32\textwidth]{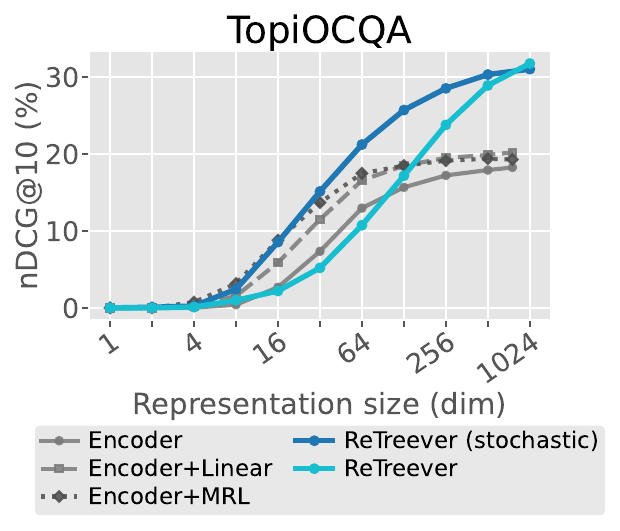} &
\includegraphics[width=0.32\textwidth]{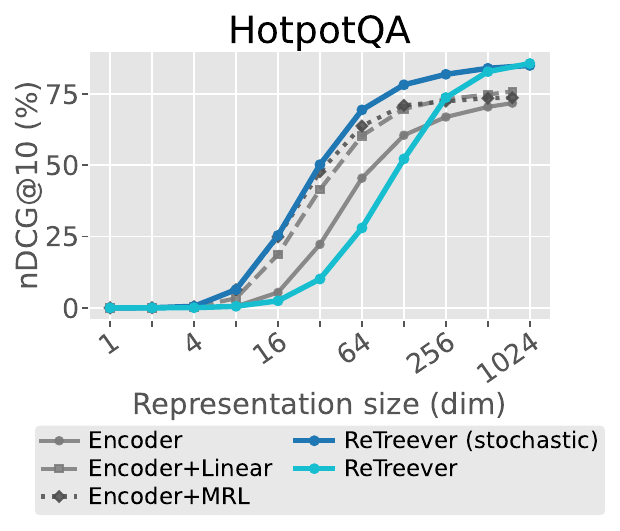} \\
\includegraphics[width=0.32\textwidth]{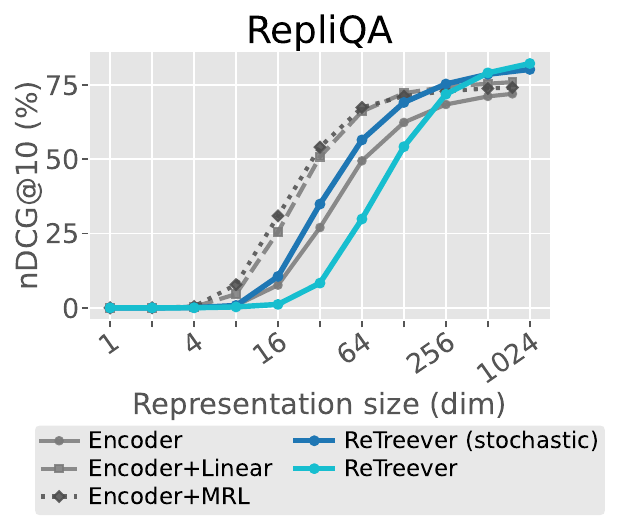} &
\includegraphics[width=0.32\textwidth]{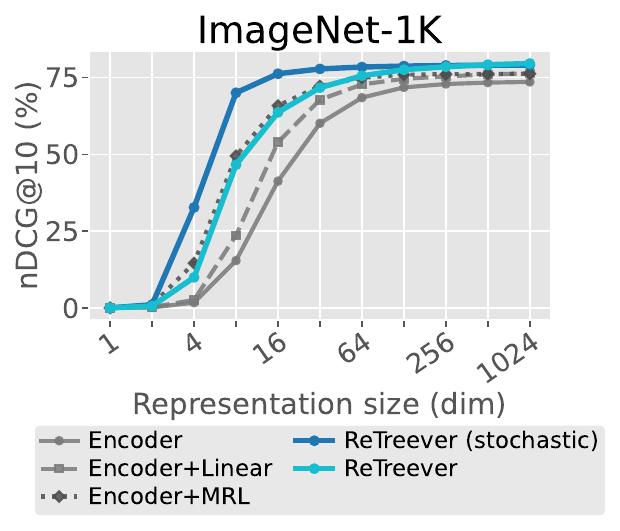} &
\includegraphics[width=0.32\textwidth]{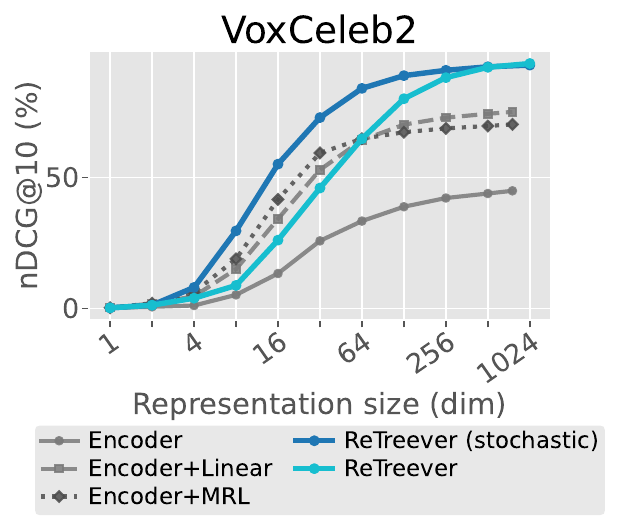}
\end{tabular}

\caption{Coarse-to-Fine Representations Across Datasets}
\label{fig:all_dataset_level_wise}
\end{figure}

\Cref{fig:all_dataset_level_wise} compares coarse-to-fine representation quality by reporting \textsc{nDCG@10} as a function of representation size. Shallower tree levels yield shorter, coarser representations, enabling faster retrieval at the cost of reduced \textsc{nDCG}, thus inducing an explicit latency–quality tradeoff. \textsc{ReTreever-Stochastic} is trained to strengthen these intermediate-level representations, and we observe this consistently across datasets: at coarse and intermediate dimensions, \textsc{ReTreever-Stochastic} outperforms \textsc{ReTreever} across nearly all sizes, while \ourmodel{} edges ahead  at the finest level, as expected when additional stochastic constraints trade a small amount of peak performance for improved coarse representations.

For baselines, we obtain coarse representations from the base encoder by truncating the embedding to its first $k$ dimensions. Since the base encoder is not tuned for our retrieval tasks, we also include an \textit{Encoder+Linear} adapter baseline, which is conceptually similar to \ourmodel{} as a lightweight adaptation layer on top of a frozen encoder. Finally, we compare against Matryoshka Representation Learning (MRL), which explicitly trains embeddings to be retrieval-effective under truncation. Across 5 of the 6 datasets shown, \textsc{ReTreever-Stochastic} provides the best \textsc{nDCG@10} on most coarse granularities, indicating that our gains come not only from explicit coarse-to-fine training (as in MRL) but also from the hierarchical tree structure that induces task-aligned semantic groupings. On \textsc{RepLiQA}, we observe a different behavior: the probability-bounding constraints can hurt performance at the coarsest levels, whereas relaxing them improves coarse representations; we analyze this effect further in \cref{app:pcl_vs_pp}.

\section{Multi-index Performance Across Datasets}

As described in \Cref{sec:inference_transparency_latency}, we consider two indexing strategies for \ourmodel{}. 
In \textit{single-index retrieval} (see \Cref{app:c2f_rep}), the probability vector produced by routing a text item through the tree to a given tree level is treated as a dense representation. We build a single FAISS-style index over these representations for the entire corpus, and retrieve by mapping queries into the same representation space and performing nearest-neighbor search.

In \textit{multi-index retrieval}, we instead exploit the fact that \ourmodel{} outputs a distribution over tree nodes, which can be interpreted as a distribution over latent, hierarchical concepts induced by the learned tree. We use these latent concepts to cluster documents and restrict search to only the most relevant clusters. Concretely, at level $10$ of a depth-$10$ tree there are $2^{10}=1024$ leaf nodes; we assign each document to the leaf with maximum assignment probability (i.e., the \emph{argmax} of its leaf distribution), and build a separate index per leaf cluster. At query time, we route the query through the same tree, select its top-$k$ leaf assignments, and search only the corresponding indices. This can substantially reduce the number of documents examined, with parameter $k$ allowing us trade off accuracy for latency.

We report \textit{data access percentage} as a proxy for latency, since retrieval time is approximately linear in the number of documents searched in this setup. This clustering-based retrieval is conceptually related to standard approximate search methods such as IVF and ScaNN, which we include as baselines.

\begin{figure}[ht]
\centering
\begin{tabular}{ccc}
\includegraphics[width=0.32\textwidth]{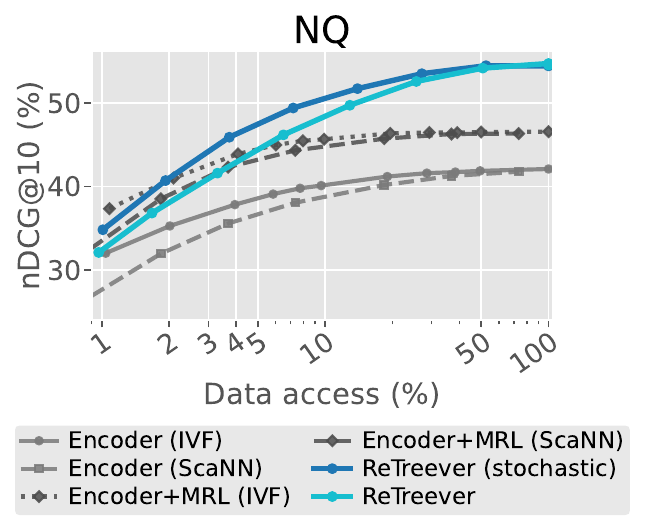} &
\includegraphics[width=0.32\textwidth]{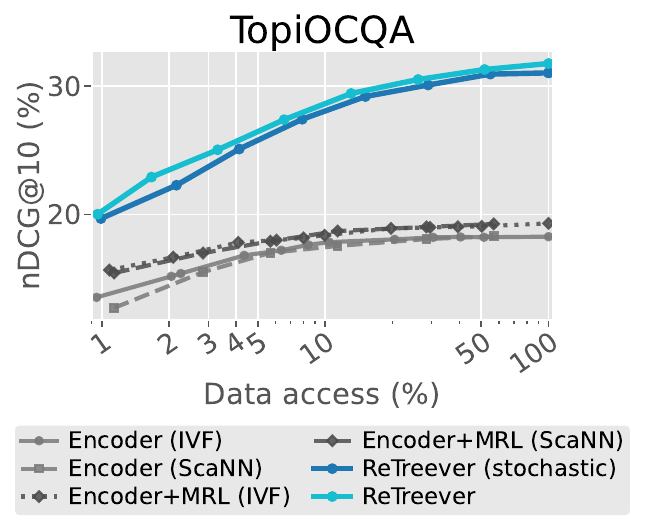} &
\includegraphics[width=0.32\textwidth]{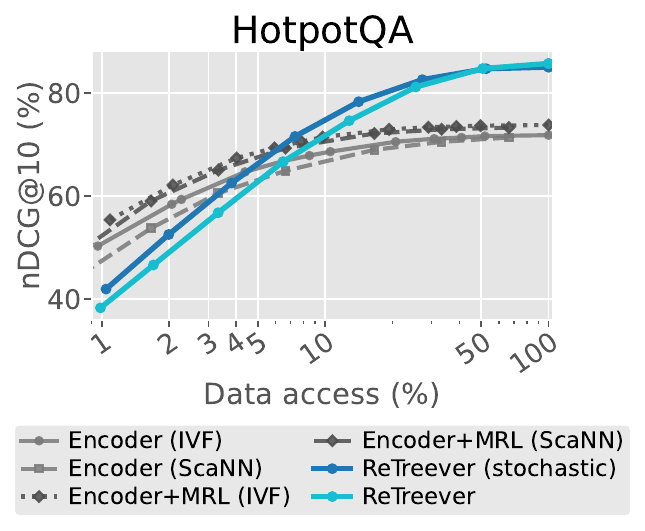} \\
\includegraphics[width=0.32\textwidth]{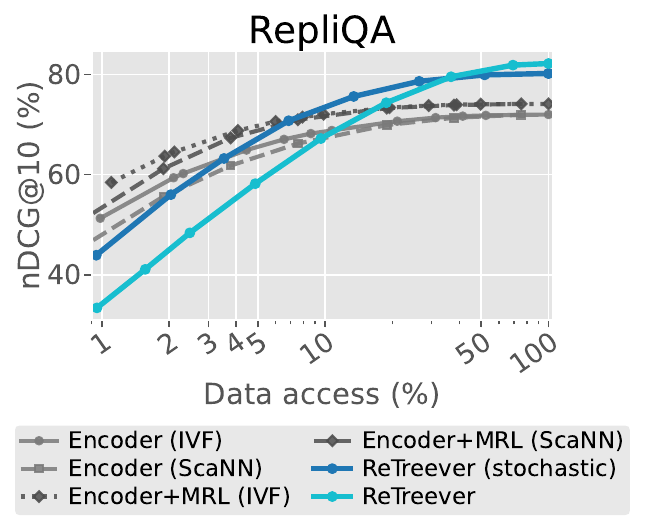} &
\includegraphics[width=0.32\textwidth]{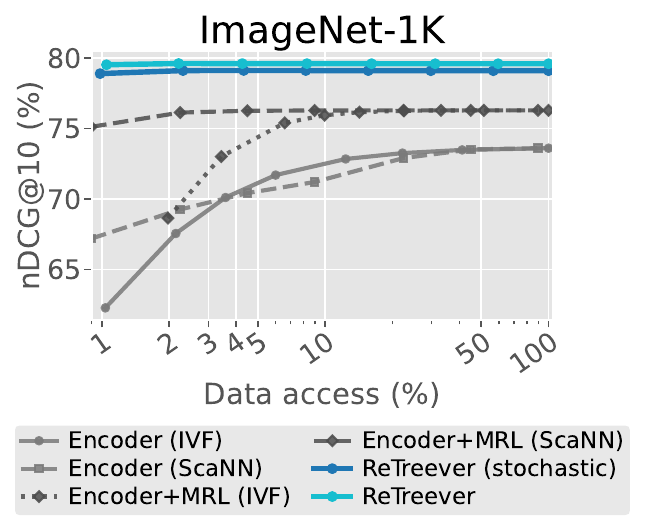} &
\includegraphics[width=0.32\textwidth]{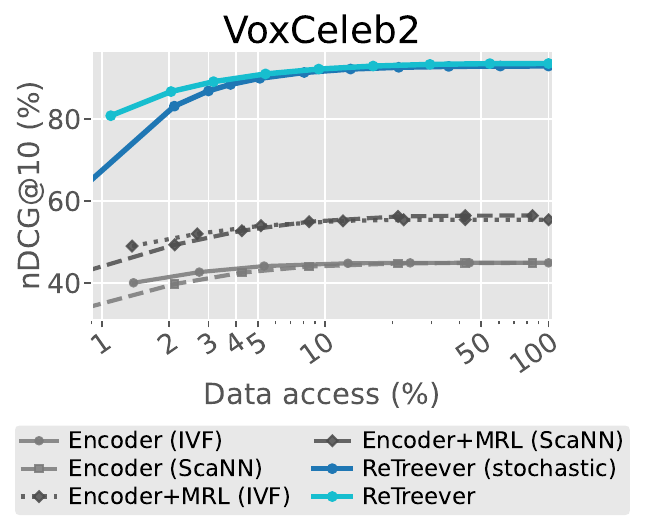}
\end{tabular}
\caption{Multi index Results over all datasets}
\label{fig:all_dataset_multi_index_results}
\end{figure}

\Cref{fig:all_dataset_multi_index_results} compares these approaches across six datasets. We evaluate IVF and ScaNN built on the base encoder representations. Since the base encoder is not tuned to these retrieval tasks, we also include IVF and ScaNN on top of an \textsc{Encoder+MRL} baseline, which explicitly trains coarse-to-fine representations and is therefore the closest non-tree analogue to our setting.

Overall, \ourmodel{} provides a favorable quality--latency tradeoff. On \textsc{TopiOCQA}, \textsc{VoxCeleb2}, and \textsc{ImageNet1K}, \ourmodel{} outperforms the approximate-search baselines at every access budget, and on \textsc{NQ} it is better at nearly all budgets. On \textsc{HotpotQA} and \textsc{RepLiQA}, \ourmodel{} trails the baselines at very small access budgets, but becomes competitive around the $5\%$ and $10\%$ access points respectively, still offering substantial latency reductions (often by more than an order of magnitude) with limited loss in retrieval quality.

We hypothesize that \textsc{HotpotQA} and \textsc{RepLiQA} induce more diffuse concept structure, requiring a broader set of leaves to cover the relevant evidence. This is particularly plausible for \textsc{HotpotQA}, where multiple supporting passages are combined into a single positive context, increasing the number of distinct concepts that must be retrieved jointly. In such cases, deeper trees (i.e., more leaves) should allow finer partitions and improve the multi-index tradeoff, which we leave as a promising direction for future work.

We further hypothesize that \textsc{TopiOCQA}, \textsc{ImageNet1K}, and \textsc{VoxCeleb2} yield sharper hierarchical organization because their queries are more informative and constrain the relevant evidence more strongly. In \textsc{TopiOCQA}, queries include conversational history, which typically disambiguates entities and intent and reduces the set of plausible matching contexts. In \textsc{ImageNet1K} and \textsc{VoxCeleb2}, queries and contexts are drawn from the same modality and distribution (image-to-image and audio-to-audio retrieval, respectively), so relevance is expressed through aligned low-level and mid-level features rather than a cross-distribution semantic mapping. In these settings, the tree can learn high-purity partitions (see \cref{fig:nmi_vs_level}), concentrating probability mass on a small number of leaves and making multi-index retrieval efficient.

\section{DistilBERT Over All Datasets}

\Cref{tab:full-rep-app} reports \textsc{Recall@10} and \textsc{nDCG@10} on the four text retrieval datasets (\textsc{NQ}, \textsc{HotpotQA}, \textsc{RepLiQA}, and \textsc{TopiOCQA}) using \texttt{sentence-transformers/msmarco-distilbert-cos-v5} as the base encoder. We evaluate finest-level retrieval using (i) the base encoder’s dense representations and (ii) \ourmodel{}’s leaf assignment distributions treated as representations, for both \textsc{ReTreever} and \textsc{ReTreever-Stochastic}.

Since the base encoder is not tuned to these downstream datasets, we additionally compare against adapter baselines that fine-tune the encoder with either a linear projection (\textsc{Encoder+Linear}) or a two-layer MLP with an up-projection and down-projection with nonlinearity and dropout (\textsc{Encoder+MLP}). Finally, we include \textsc{Encoder+MRL} (Matryoshka Representation Learning), which explicitly trains coarse-to-fine representations and is therefore the closest non-tree analogue to our setting.

Across all datasets, adapting the encoder with a linear or MLP head consistently improves both \textsc{Recall@10} and \textsc{nDCG@10} over the base encoder. In contrast, \textsc{Encoder+MRL} yields weaker finest-level performance, consistent with the general tradeoff that enforcing coarse-to-fine structure can reduce peak performance at the finest granularity. We observe a similar pattern for \ourmodel{}: at the finest level, \textsc{ReTreever} is consistently stronger than \textsc{ReTreever-Stochastic}, while \textsc{ReTreever-Stochastic} provides substantially better coarse representations across levels (see \Cref{fig:all_dataset_level_wise}). Overall, both \ourmodel{} variants outperform all baselines on all datasets under the finest-level \textsc{Recall@10} and \textsc{nDCG@10} metrics in \Cref{tab:full-rep-app}.

\begin{table}[H]
\centering
\vspace{10pt}
\caption{Full representation retrieval \textsc{Recall@10} (\%) and \textsc{nDCG@10} (\%), with best results per metric and dataset in bold, on the text retrieval datasets.}
\label{tab:full-rep-app}
\vspace{-10pt}
\begin{center}
\begin{sc}
\begin{tabular}{l@{\hskip 4pt}c@{\hskip 4pt}c@{\hskip 4pt}c@{\hskip 4pt}c@{\hskip 4pt}c@{\hskip 4pt}c@{\hskip 4pt}c@{\hskip 4pt}c}
\toprule
\textbf{Model} & \multicolumn{2}{c}{\textsc{NQ}} & \multicolumn{2}{c}{\textsc{HotpotQA}} & \multicolumn{2}{c}{\textsc{TopiOCQA}} & \multicolumn{2}{c}{\textsc{RepLiQA}} \\
& \scriptsize{Rec${@}10$ $\uparrow$} & \scriptsize{nDCG${@}10$ $\uparrow$} 
& \scriptsize{Rec${@}10$ $\uparrow$} & \scriptsize{nDCG${@}10$ $\uparrow$}
& \scriptsize{Rec${@}10$ $\uparrow$} & \scriptsize{nDCG${@}10$ $\uparrow$}
& \scriptsize{Rec${@}10$ $\uparrow$} & \scriptsize{nDCG${@}10$ $\uparrow$} \\ 
\midrule
Encoder & 57.53 & 42.15 & 82.14 & 71.82 & 31.03 & 18.26 & 82.39 & 72.02 \\
Encoder + Linear & 68.61 & 47.62 & 86.51 & 75.90 & 36.86 & 20.20 & 86.50 & 75.96 \\
Encoder + MLP & 75.10 & 52.23 & 87.72 & 76.88 & 36.24 & 20.63 & 85.66 & 74.99 \\
Encoder + MRL & 67.10 & 46.60 & 84.46 & 73.70 & 34.50 & 19.30 & 85.20 & 74.15 \\
\midrule
Retreever - Stochastic & 77.36 & 54.48 & 92.95 & 85.04 & 51.80 & 31.00 & 89.36 & 80.23 \\
Retreever & \textbf{77.50} & \textbf{54.58} & \textbf{93.47} & \textbf{85.71} & \textbf{52.11} & \textbf{31.78} & \textbf{90.84} & \textbf{82.22} \\
\bottomrule
\end{tabular}
\end{sc}
\end{center}
\vskip0.5cm
\end{table}

\section{Comparing Different Text Encoders}
\label{app:different_text_encoders}

We repeat the same set of baselines and training variants as in \Cref{tab:full-rep-app}, but replace the base encoder with \texttt{BAAI/bge-large-en-v1.5}. Results are reported in \Cref{tab:bge-fullrep-ndcg10-pct}. BGE is a larger encoder (roughly 600M parameters) that is pre-trained with masked language modeling on large-scale text and then fine-tuned on retrieval and QA-style tasks.

A notable difference relative to DistilBERT is that, when the encoder is already strongly fine-tuned for question-answering, additional adapter-style fine-tuning can be less reliable and may even degrade performance. This effect is visible in \Cref{tab:bge-fullrep-ndcg10-pct}: adding an MLP adapter decreases performance on \textsc{RepLiQA} relative to the base encoder, and on \textsc{HotpotQA} a linear adapter outperforms MLP. The same sensitivity is reflected in \ourmodel{} results, suggesting that when the backbone already provides a highly specialized representation space, further adaptation must be applied more cautiously.

In contrast, \textsc{TopiOCQA} is conversational and likely less aligned with BGE’s fine-tuning distribution, where we observe a more intuitive improvement pattern under adaptation. Importantly, regardless of the absolute finest-level performance, \ourmodel{} retains its key advantages: it provides a controllable latency--quality tradeoff via single-index and multi-index retrieval, and it induces an explicit hierarchical organization of the corpus that supports interpretability and efficient search, independent of the choice of base encoder.

\begin{table}[H]
\centering
\vspace{10pt}
\caption{Full-representation retrieval \textsc{nDCG@10} (\%) with a BGE encoder across text retrieval datasets, comparing the base encoder, adapter baselines, and \ourmodel{} variants. Best results per dataset are in bold.}
\label{tab:bge-fullrep-ndcg10-pct}
\begin{center}
\begin{sc}
\begin{tabular}{l@{\hskip 6pt}c@{\hskip 6pt}c@{\hskip 6pt}c@{\hskip 6pt}c}
\toprule
\textbf{Model} & \multicolumn{1}{c}{\textsc{NQ}} & \multicolumn{1}{c}{\textsc{HotpotQA}} & \multicolumn{1}{c}{\textsc{TopiOCQA}} & \multicolumn{1}{c}{\textsc{RepLiQA}} \\
& \scriptsize{nDCG${@}10$ $\uparrow$} & \scriptsize{nDCG${@}10$ $\uparrow$} & \scriptsize{nDCG${@}10$ $\uparrow$} & \scriptsize{nDCG${@}10$ $\uparrow$} \\
\midrule
Encoder & 55.57 & 91.54 & 22.19 & 86.42 \\
Encoder + MLP & 57.20 & 91.67 & 25.02 & 84.68 \\
Encoder + Linear & 58.19 & \textbf{91.98} & 24.91 & \textbf{87.01} \\
Encoder + MRL & 57.05 & 90.40 & 24.35 & 85.47 \\
\midrule
Retreever + Stochastic & 60.26 & 90.06 & \textbf{30.11} & 84.45 \\
Retreever & \textbf{60.46} & 91.05 & 27.82 & 85.74 \\
\bottomrule
\end{tabular}
\end{sc}
\end{center}
\end{table}

\section{Comparing Different Image Encoders}
\label{app:compare_image_encoders}

We evaluate \ourmodel{} with three image encoders spanning a wide range of capacity and inductive bias: (i) \textbf{ResNet-50}, a convolutional backbone with roughly $25.6$M parameters and a $2048$-dimensional feature representation; 
(ii) \textbf{DINOv2-Large} (\texttt{ViT-L/14}), a self-supervised Vision Transformer with a $1024$-dimensional embedding and roughly $300$M parameters; and finally (iii) \textbf{CLIP ViT-L/14} (\texttt{openai/clip-vit-large-patch14}), a vision-language model whose vision tower uses width $1024$ and projects to a $768$-dimensional shared embedding space (model size $\approx 427$M parameters).

Across all three encoders, we consistently observe that \ourmodel{} yields markedly stronger  representations at both coarse and fine levels, as compared to the non-tree baselines, including Matryoshka Representation Learning (MRL), which explicitly targets coarse-to-fine behavior but does not impose a hierarchical routing structure. As expected, the \textsc{ReTreever-Stochastic} variant produces the strongest coarse representations, as its explicitly trained to do so. Overall, these results support the view that learning a retrieval-optimized tree induces semantically meaningful hierarchical organization of the corpus, which directly translates into improved retrieval quality at various representation levels.
\vspace{10pt}
\begin{figure}[H]
    \centering
    \includegraphics[width=\textwidth]{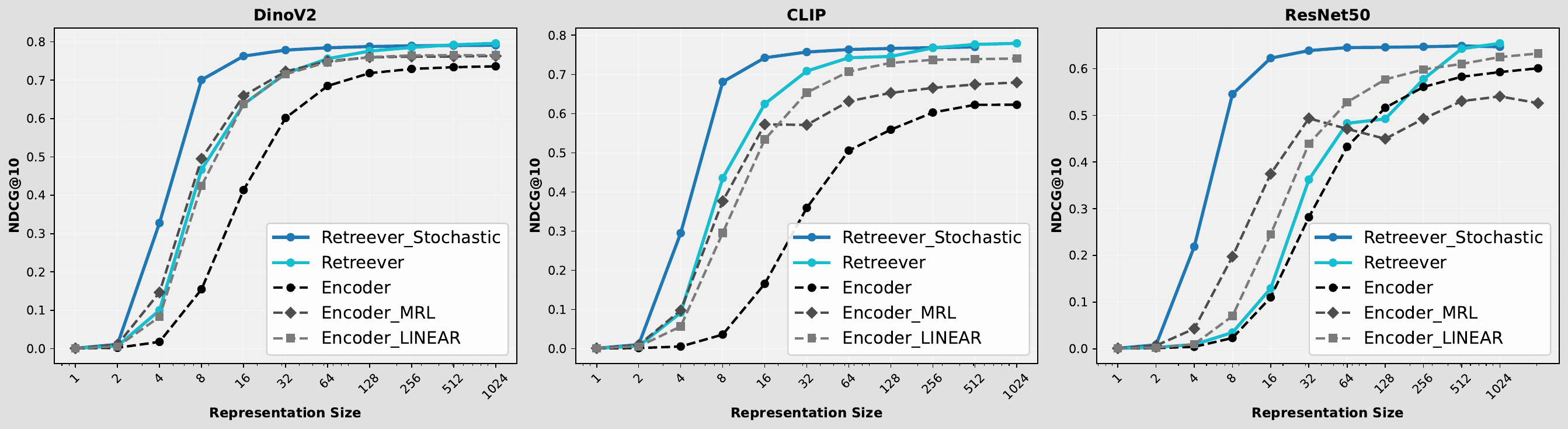}
    \caption{Level wise retrieval performance for three image encoders}
    \label{fig:image_encoders}
\end{figure}

\section{Retreever Latency vs Representation Size}
\label{app:rep_size_vs_latency}

\begin{figure}[H]
    \centering
    \begin{minipage}{0.49\textwidth}
        \centering
        \includegraphics[width=\textwidth]{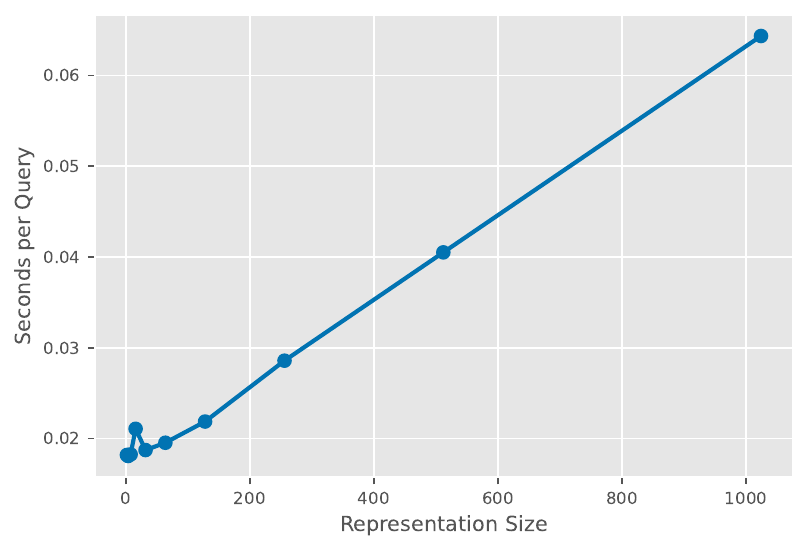}
    \end{minipage}
    \hfill
    \begin{minipage}{0.49\textwidth}
        \centering
        \includegraphics[width=\textwidth]{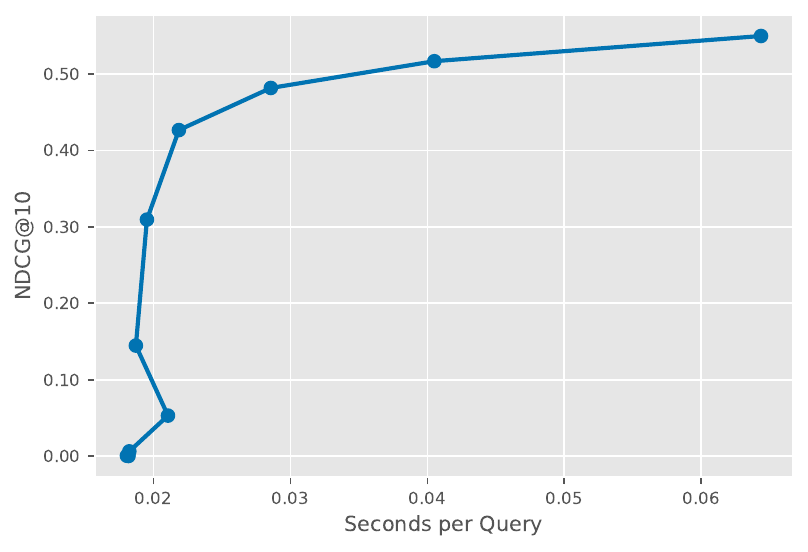}
        \label{fig:latency_vs_ndcg}
    \end{minipage}
    \caption{Trade-off between retrieval speed and effectiveness in Retreever. The left plot shows a linear dependence between the representation size from \ourmodel{} and the query latency, while the right plot demonstrates that a speedup in query inference with only a slight reduction in retrieval performance.}
    \label{fig:latency_tradeoff}
\end{figure}

We analyze the relationship between embedding size obtained from \ourmodel{} and its impact on inference latency and retrieval performance. As shown in \autoref{fig:latency_tradeoff}, the seconds per query (\texttt{sec/query}) increase as the representation size grows. This trend is expected, as larger representations require more computation and memory access during retrieval. We observe, that the inference latency is linearly related to the size of the representation obtained from \ourmodel{}. We also note from the figure that its possible to obtain a significant decrease in query latency without sacrificing on retrieval accuracy by much.

\autoref{fig:latency_tradeoff} further illustrates the trade-off between retrieval effectiveness, measured by NDCG@10, and query latency. While larger representations significantly increase model latency, they offer only marginal improvements in retrieval performance. This provides a way to balance query latency and retrieval performance by selecting an appropriate embedding size from \ourmodel{} to achieve the desired trade-off.

\begin{figure*}
    \centering
    \includegraphics[width=\textwidth]{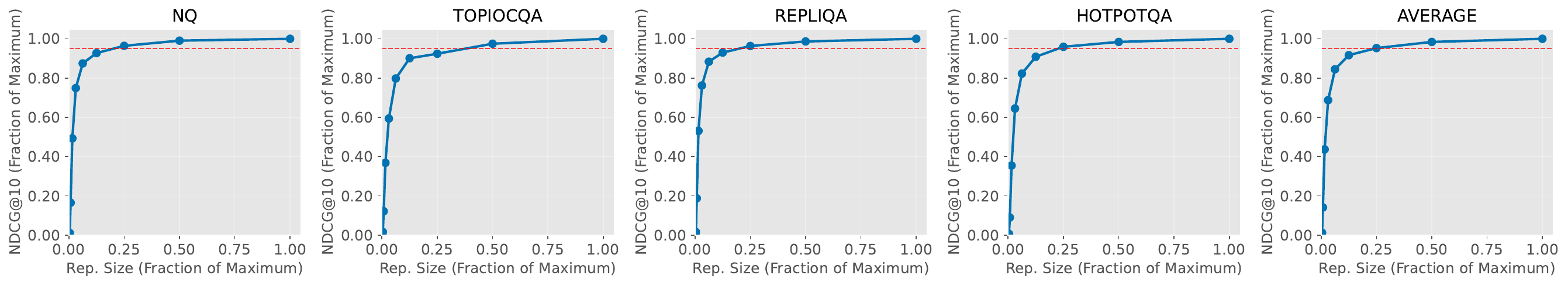}
    \includegraphics[width=\textwidth]{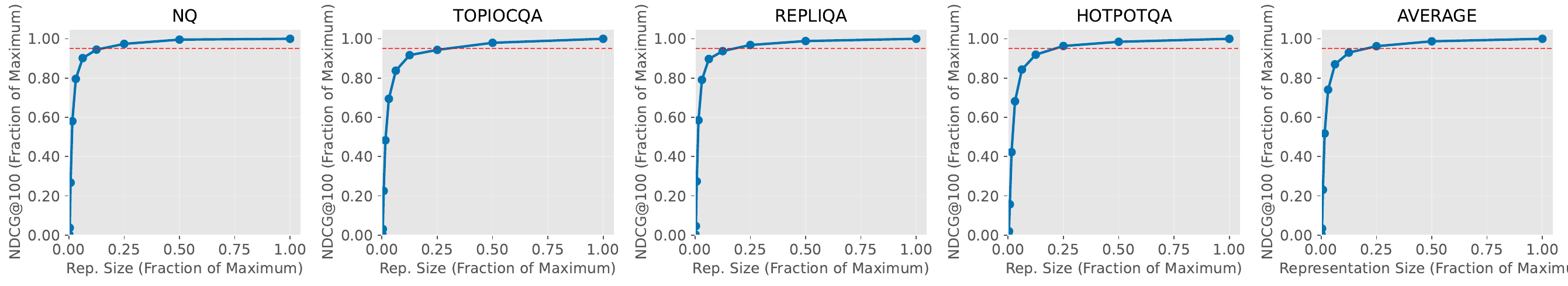}
    \caption{\textsc{NDCG@10} (above) and \textsc{NDCG@100} (below)  performance vs. representation size trade-off for \textsc{ReTreever-Stochastic}. All values are normalized as fractions of their respective maximums. Dashed red lines show 95\% performance thresholds. The average across datasets (rightmost) shows that 95\% of peak performance is maintained even at 1/4 maximum representation size (i.e. dimension size 256) for \textsc{nDCG@10} and at 1/8 maximum representation size (i.e. dimension size 128) for \textsc{nDCG@100}. }
    \label{fig:latency_perf_tradeoff_ndcg10}
\end{figure*}

\section{Quantitative Metrics for Interpretability}
\label{app:quant_metrics}

\begin{figure}[H]
\centering
\begin{tabular}{ccc}
\includegraphics[width=0.45\textwidth]{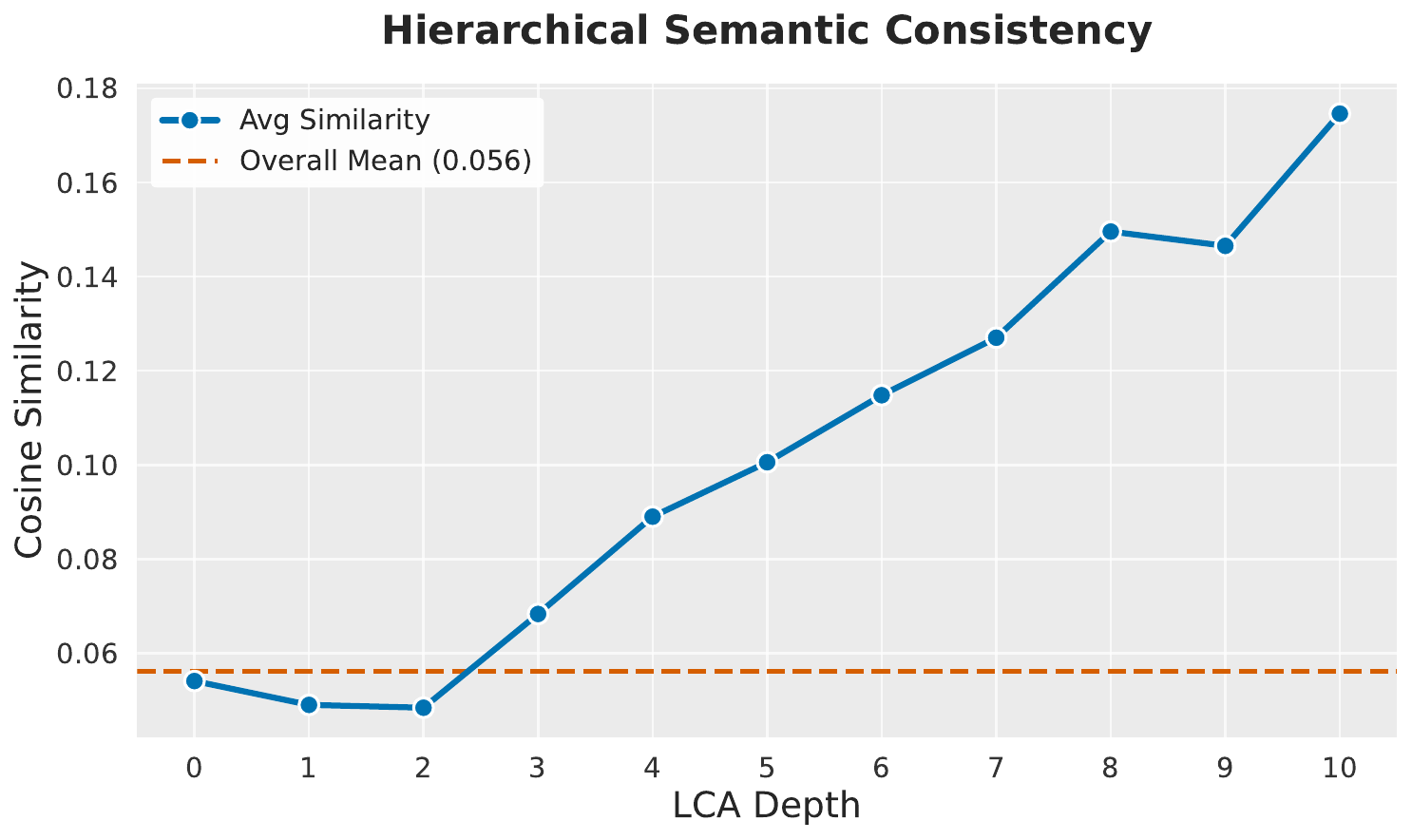} &
\includegraphics[width=0.45\textwidth]{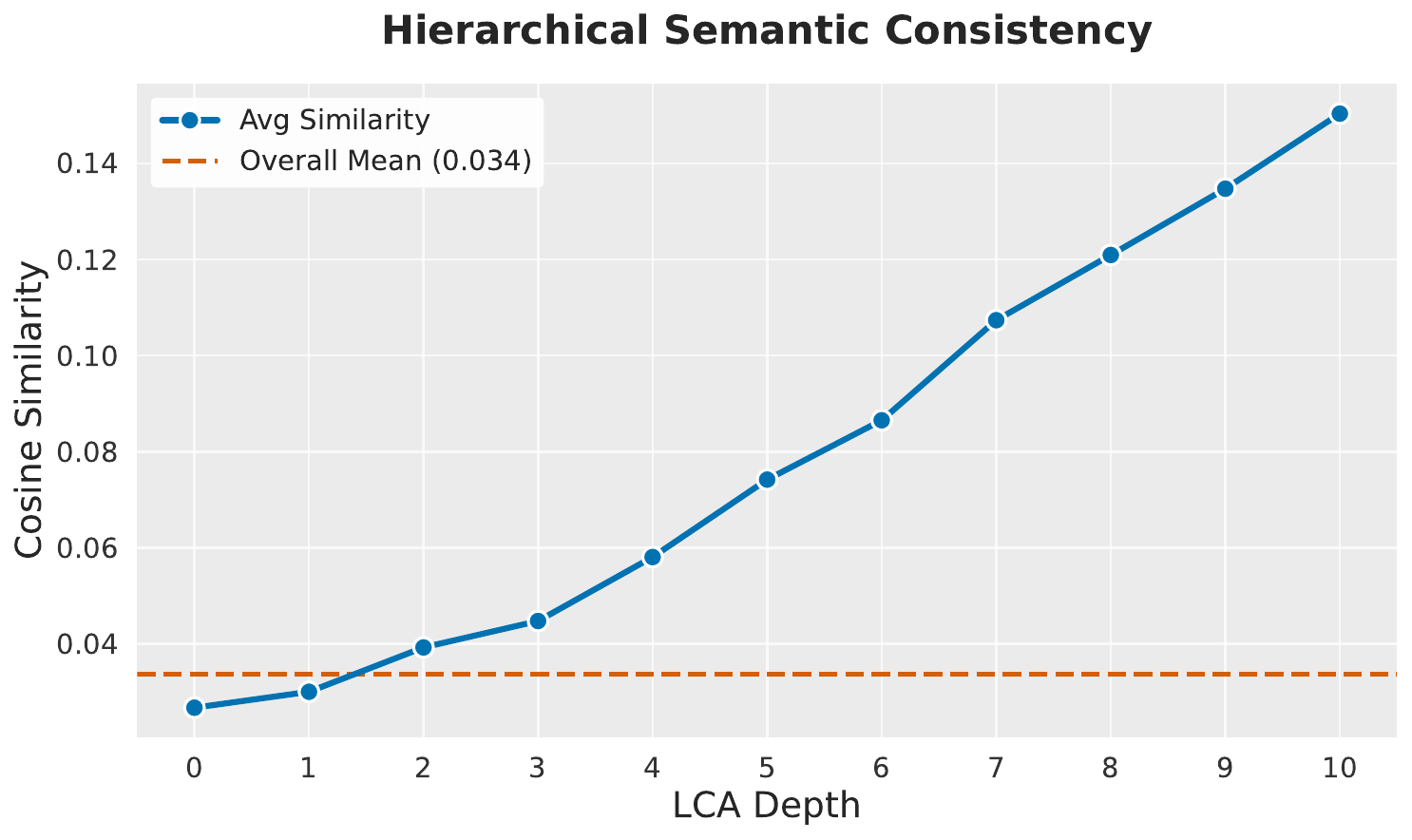} \\
\includegraphics[width=0.45\textwidth]{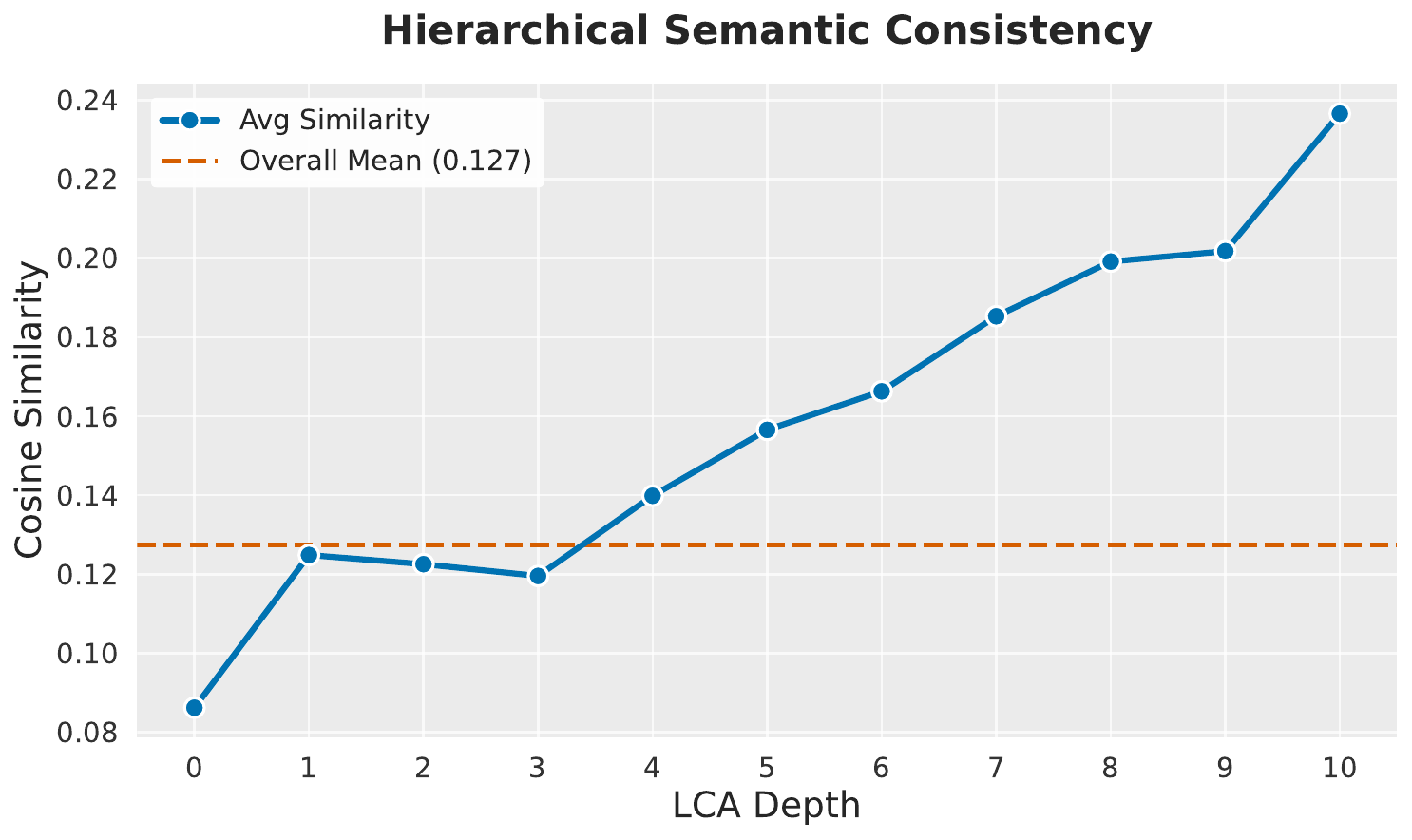} &
\includegraphics[width=0.45\textwidth]{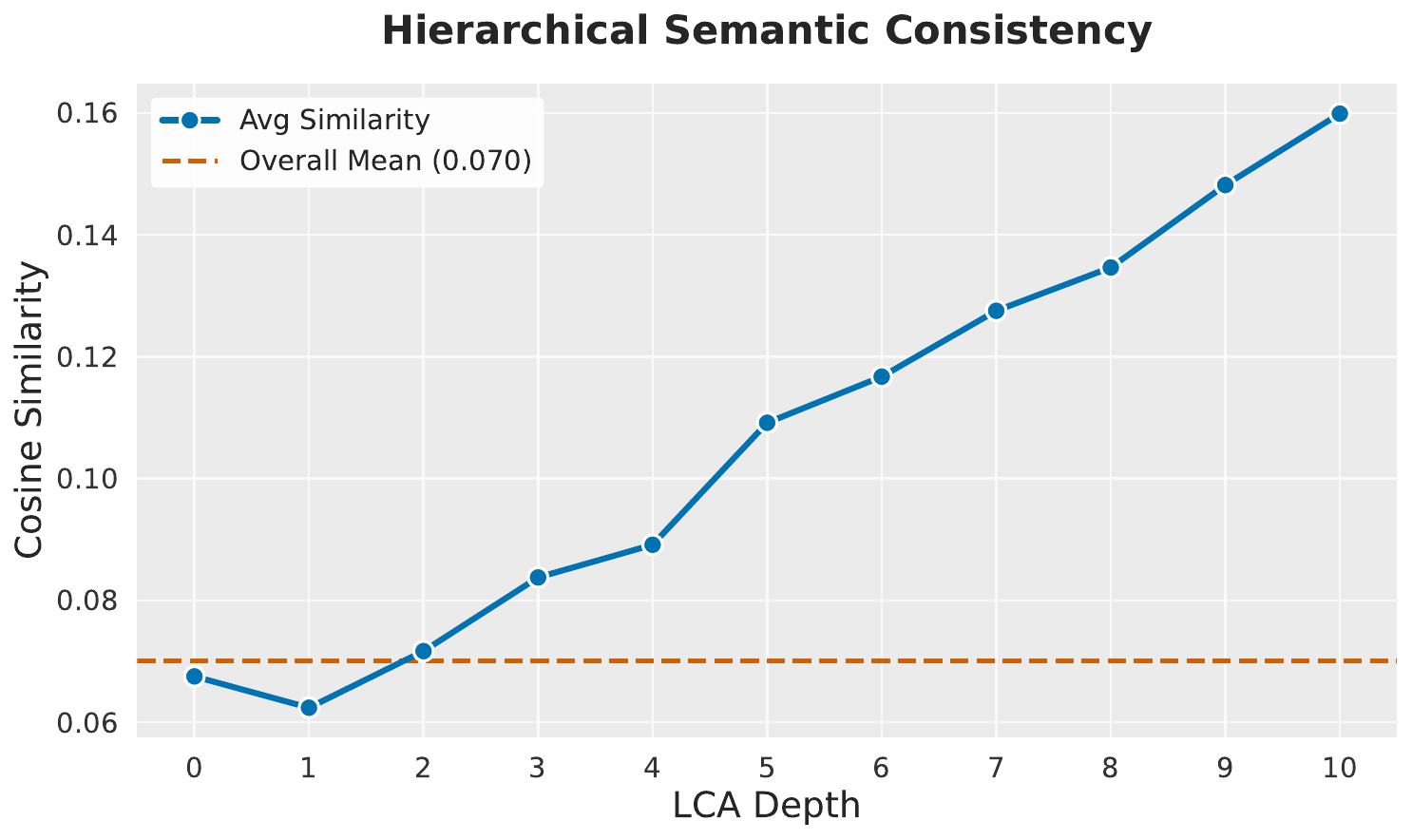} 
\end{tabular}
\caption{Cosine similarity vs LCA depth for text pairs}
\label{fig:all_dataset_hier_analysis}
\end{figure}

We evaluate whether ReTreever learns a meaningful refinement hierarchy by measuring hierarchical semantic consistency. For randomly sampled pairs of documents, we assign each document to its most probable leaf (argmax) and compute the depth of the lowest common ancestor (LCA) of the two leaves. We then compute cosine similarity between the documents’ original encoder embeddings and report the average similarity as a function of LCA depth. Across all four text datasets (see \cref{fig:all_dataset_hier_analysis}, similarity increases monotonically with LCA depth, and is substantially above the global mean similarity baseline for deep LCAs, indicating that deeper regions of the tree correspond to semantically tighter groupings. This supports the claim that the learned tree induces a coherent hierarchy with increasing semantic refinement toward deeper levels. 

\paragraph{Class-Node Alignment Across Tree Levels (ImageNet-1K).}
To assess whether the learned tree induces semantically meaningful partitions on ImageNet-1K, we measure how well node assignments at each depth align with ground-truth class labels.
For each tree level $\ell \in \{1,\dots,D\}$, we extract the model's assignment scores over the $2^\ell$ nodes at that level (a contiguous slice of the full node-score vector), convert the soft assignments into a hard node index via $\arg\max$, and compute the \emph{Normalized Mutual Information} (NMI) between the resulting node assignments and the ImageNet class labels.
NMI is an information-theoretic measure of agreement between two clusterings, normalized to lie in $[0,1]$, where $0$ indicates independence and $1$ indicates perfect correspondence.

\begin{figure}[H]
    \centering
    \includegraphics[width=0.6 \textwidth]{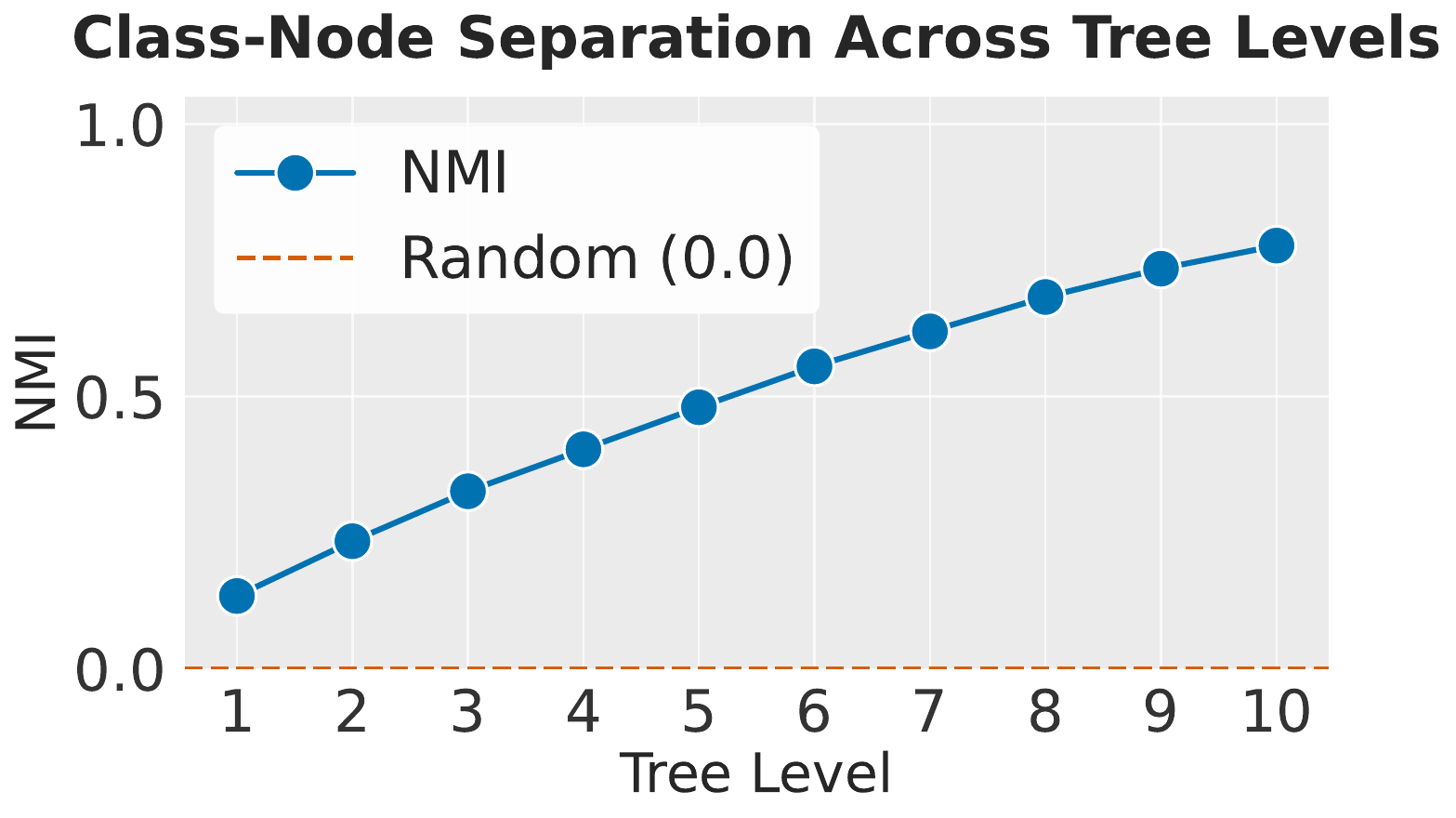}
    \caption{NMI between tree node assignments and imagenet1K classes increase with depth. Baseline is no correlation between the two having an NMI of 0.0}
    \label{fig:nmi_vs_level}
\end{figure}

An increasing NMI with depth indicates that deeper levels of the tree progressively refine the partition of the dataset into more class-consistent groups, as expected from a coarse-to-fine hierarchy.
In our experiments, NMI rises consistently with tree depth and approaches $\approx 0.8$ at the deepest level, suggesting that the learned hierarchy strongly separates ImageNet classes into distinct regions of the tree rather than producing arbitrary splits. \cref{fig:nmi_vs_level}
This supports the claim that ReTreever's routing structure captures meaningful semantic organization that becomes more discriminative as depth increases.

\vspace{10pt}

\section{Comparison of Different Propagation Methods}
\label{app:pcl_vs_pp}
\paragraph{Product Propagation.}
As introduced in \Cref{sec:ret_arch}, \ourmodel{} uses \emph{product propagation} to couple node-wise routing scores through the geometry of the tree. 
The split function at each branching node produces a local routing logit, and applying a sigmoid yields a local branching probability. 
However, these local decisions must be consistent with the fact that reaching a node requires taking all routing decisions along the unique path from the root. 
Product propagation enforces this constraint by defining the probability of reaching node $t$ as the product of the relevant routing probabilities along its root-to-$t$ path. 
Concretely, given a node $t$ and its ancestors $\mathcal{A}_t$ (the nodes along the path from the root to $t$), the probability of an input $\xb_i$ reaching $t$ is
\begin{equation}
T(\xb_i)_t \;=\; \zb_{i,t}\prod_{a\in\mathcal{A}_t}\zb_{i,a},
\end{equation}
where $\zb_{i,u}\in[0,1]$ denotes the (sigmoid) routing probability produced at node $u$ for input $i$.
This propagation rule yields a globally consistent distribution over nodes, and in particular a leaf-assignment distribution obtained by restricting $T(\xb_i)$ to the leaf set.

\paragraph{Path-Consistency Propagation (Relaxed Propagation).}
While product propagation enforces a strict multiplicative constraint, we found that coarse representations can benefit from a more relaxed coupling of node probabilities. 
We therefore consider a relaxed alternative in which node scores are no longer treated as independent. 
Instead, for each node $t$, we rescore it using a learned function of the raw split logits along its path to the root.
Let $r_{i,u}\in\mathbb{R}$ denote the raw split logit produced at node $u$ for input $i$. 
We define a path-conditioned rescoring function $f_{\phi}$ (implemented as a small MLP) that maps the sequence of logits on the root-to-$t$ path to a new node logit:
\begin{equation}
\tilde{r}_{i,t} \;=\; f_{\phi}\!\big(\{r_{i,u}\}_{u\in\mathcal{A}_t\cup\{t\}}\big),
\qquad
p_{i,t} \;=\; \sigma(\tilde{r}_{i,t}),
\end{equation}
where $p_{i,t}\in[0,1]$ is now interpreted directly as the probability of \emph{reaching} node $t$ (rather than a left/right branching probability at $t$). 
This introduces explicit dependence between a node and its ancestors through $f_{\phi}$, but does not by itself enforce tree-consistent paths; we therefore add a regularizer that encourages path consistency.

\paragraph{Path-Consistency Regularization.}
Given node-reaching probabilities $\{p_{i,t}\}_{t\in\mathcal{T}}$ for input $i$, the regularizer encourages high probability mass along the path to whichever node at a target depth is most likely. 
Let $\mathcal{T}_d$ be the set of nodes at depth $d$. 
We form a soft selection over nodes at depth $d$ using a temperature $\tau>0$:
\begin{equation}
w_{i,t} \;=\; \frac{\exp(p_{i,t}/\tau)}{\sum_{u\in\mathcal{T}_d}\exp(p_{i,u}/\tau)},
\qquad t\in\mathcal{T}_d .
\end{equation}
For each candidate node $t\in\mathcal{T}_d$, we score the log-probability of its entire root-to-$t$ path, and take a weighted average under $w_{i,t}$:
\begin{equation}
\mathcal{L}_{\mathrm{pc}}^{(d)} 
\;=\;
-\frac{1}{B}\sum_{i=1}^{B}\;
\sum_{t\in\mathcal{T}_d}
w_{i,t}\;
\sum_{u\in\mathcal{A}_t\cup\{t\}}
\log\!\big(p_{i,u}+\varepsilon\big),
\label{eq:path_consistency_loss}
\end{equation}
where $B$ is the batch size and $\varepsilon$ is a small constant for numerical stability. 
As $\tau\!\to\!0$, the weights concentrate on the highest-probability node at depth $d$, recovering a hard “best-path” penalty; for larger $\tau$, the loss encourages consistency across multiple plausible paths.

\begin{figure}[H]
    \centering
    \includegraphics[width=\textwidth]{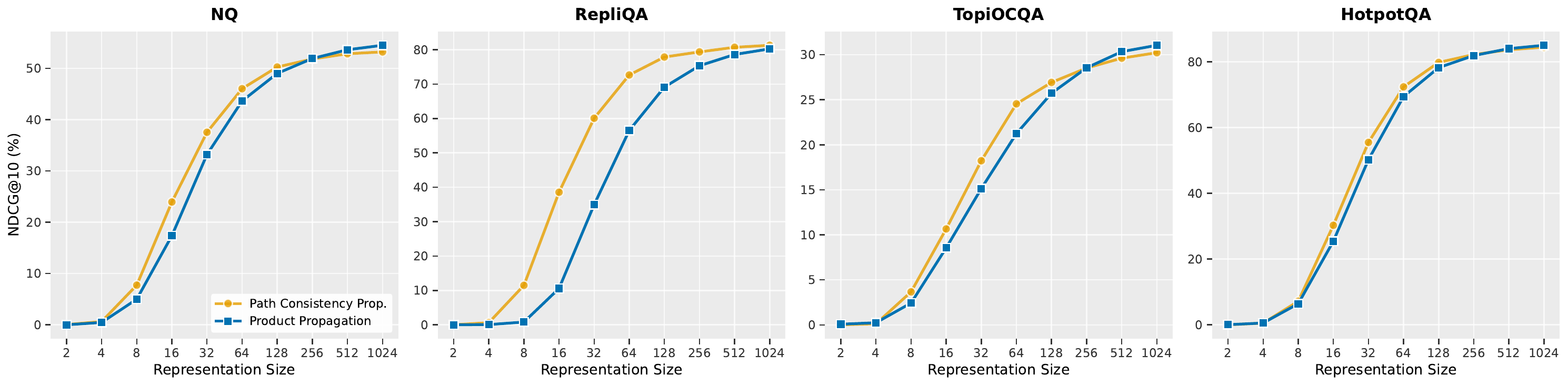}
    \caption{Level wise retrieval performance for different propagation methods}
    \label{fig:pcl_vs_pp}
\end{figure}

\paragraph{Final Objective.}
We add the path-consistency term to the main contrastive objective in \Cref{eq:constrative_loss}. 
Since we compute node probabilities for both queries and contexts, we apply the regularizer to both:
\begin{equation}
\mathcal{L}
\;=\;
\mathcal{L}_{\mathrm{InfoNCE}}
\;+\;
\lambda\Big(
\mathcal{L}_{\mathrm{pc}}(\text{queries})
+
\mathcal{L}_{\mathrm{pc}}(\text{contexts})
\Big),
\end{equation}
where $\lambda$ controls the strength of the regularization.
We compare strict product propagation against the relaxed MLP propagation with path-consistency regularization in \Cref{tab:propagation_pathconsistency_ablation}, and we see the level wise difference in \cref{fig:pcl_vs_pp}. 

We observe that while both propagation schemes achieve similar performance at the finest level, the relaxed propagation yields stronger coarse representations, most notably on RepLiQA. In particular, on RepLiQA it also outperforms the MRL baseline in \cref{app:c2f_rep} at coarse granularities. For simplicity, however, we use product propagation in all other experiments.
\vspace{10pt}

\begin{table}[H]
\centering
\vspace{10pt}
\caption{Effect of different propagation mechanisms on text retrieval performance (\textsc{nDCG@10}). Best results per dataset within each block are in bold.}
\label{tab:propagation_pathconsistency_ablation}
\vspace{-10pt}
\begin{center}
\begin{sc}
\begin{tabular}{l@{\hskip 4pt}c@{\hskip 4pt}c@{\hskip 4pt}c@{\hskip 4pt}c}
\toprule
\textbf{Model} & \textsc{NQ} & \textsc{HotpotQA} & \textsc{TopiOCQA} & \textsc{RepliQA} \\
\midrule
\multicolumn{5}{l}{\textbf{Product Propagation}} \\
\midrule
Retreever - Stochastic & 0.5448 & 0.8504 & 0.3100 & 0.8023 \\
Retreever   & \textbf{0.5458} & \textbf{0.8571} & 0.3178 & \textbf{0.8222} \\
\midrule
\multicolumn{5}{l}{\textbf{Path Consistency Propagation}} \\
\midrule
Retreever - Stochastic & 0.5417 & 0.8379 & 0.3267 & 0.8097 \\
Retreever   & 0.5374 & 0.8457 & \textbf{0.3429} & 0.8122 \\
\bottomrule
\end{tabular}
\end{sc}
\end{center}
\vskip0.5cm
\end{table}

\label{app:propagation}

\section{Interactive Tree Visualization Tool}
\label{app:viz_tool}
To facilitate the inspection and analysis of learned hierarchical structures in \textsc{ReTreever}, we developed an interactive web-based visualization tool that
renders the complete binary tree structure using Plotly.js, with automatic layout positioning based on node depth and rank. We make this tool available anonymously here: \href{https://euphonious-brigadeiros-cf5172.netlify.app/}{https://euphonious-brigadeiros-cf5172.netlify.app/}

\paragraph{Construction} In the trees representing text datasets, upon selecting any node, the tool displays comprehensive metadata including the node identifier and tree level, an automatically generated semantic label, a summary description of the assigned document cluster, and dominant keyword themes. We use the \texttt{Qwen/Qwen2.5-72B-Instruct} model to generate this semantic metadata. For leaf nodes, we prompt the model to extract the most specific, fine-grained topics shared by documents assigned to the node. For internal nodes, we prompt the model to (i) identify salient keywords from a random subset of highly ranked documents in the node’s cluster, and (ii) jointly analyze the child nodes’ summaries and keywords to infer the broader concept represented by the parent. In addition, we ask the model to characterize how the left and right subtrees differ by producing child-specific keywords that are indicative of one child (left or right) but not the other, highlighting both shared domains and discriminative themes. 

\begin{figure}[t]
\centering
\begin{tabular}{cc}
\includegraphics[width=0.45\textwidth]{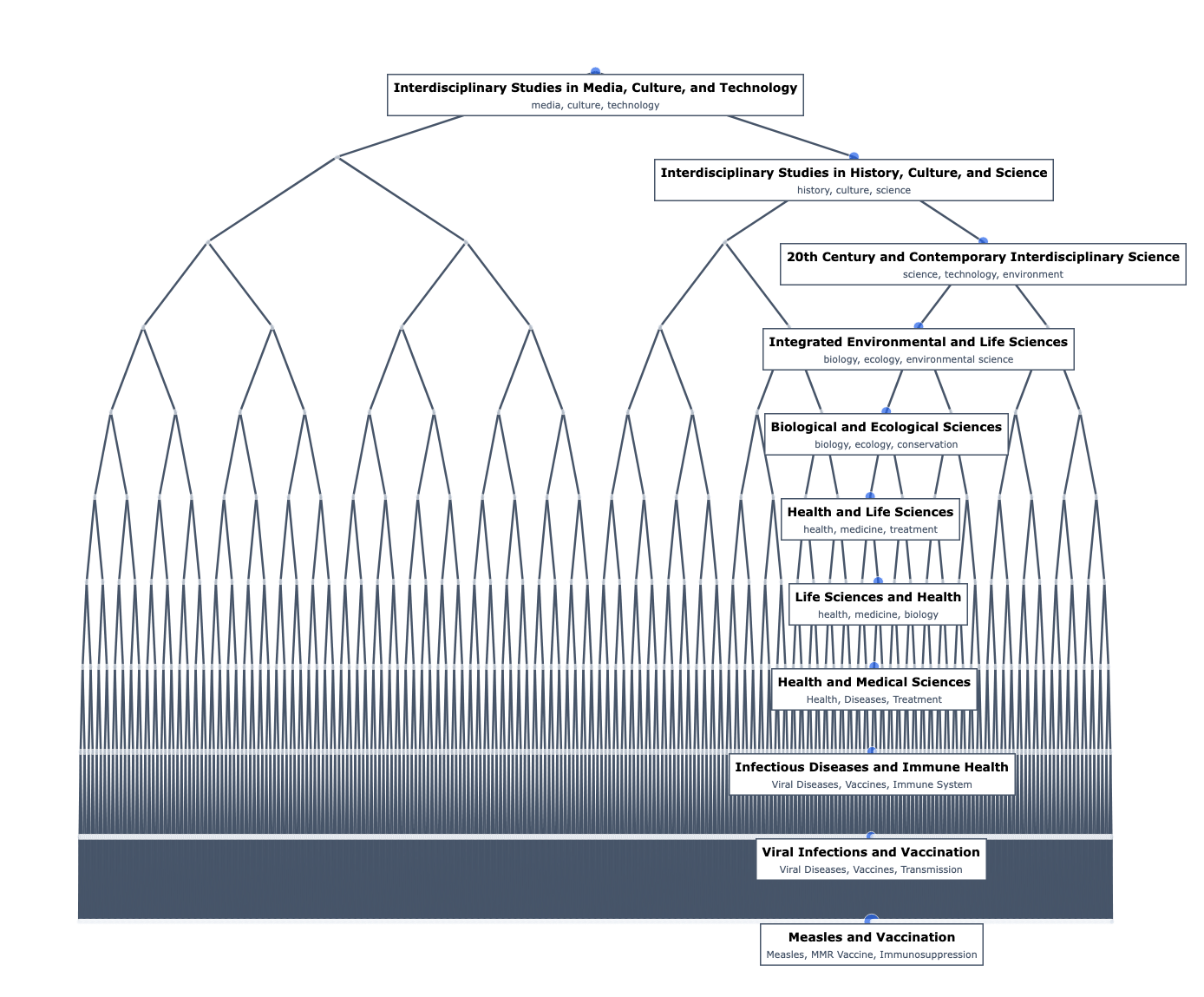} &
\includegraphics[width=0.45\textwidth]{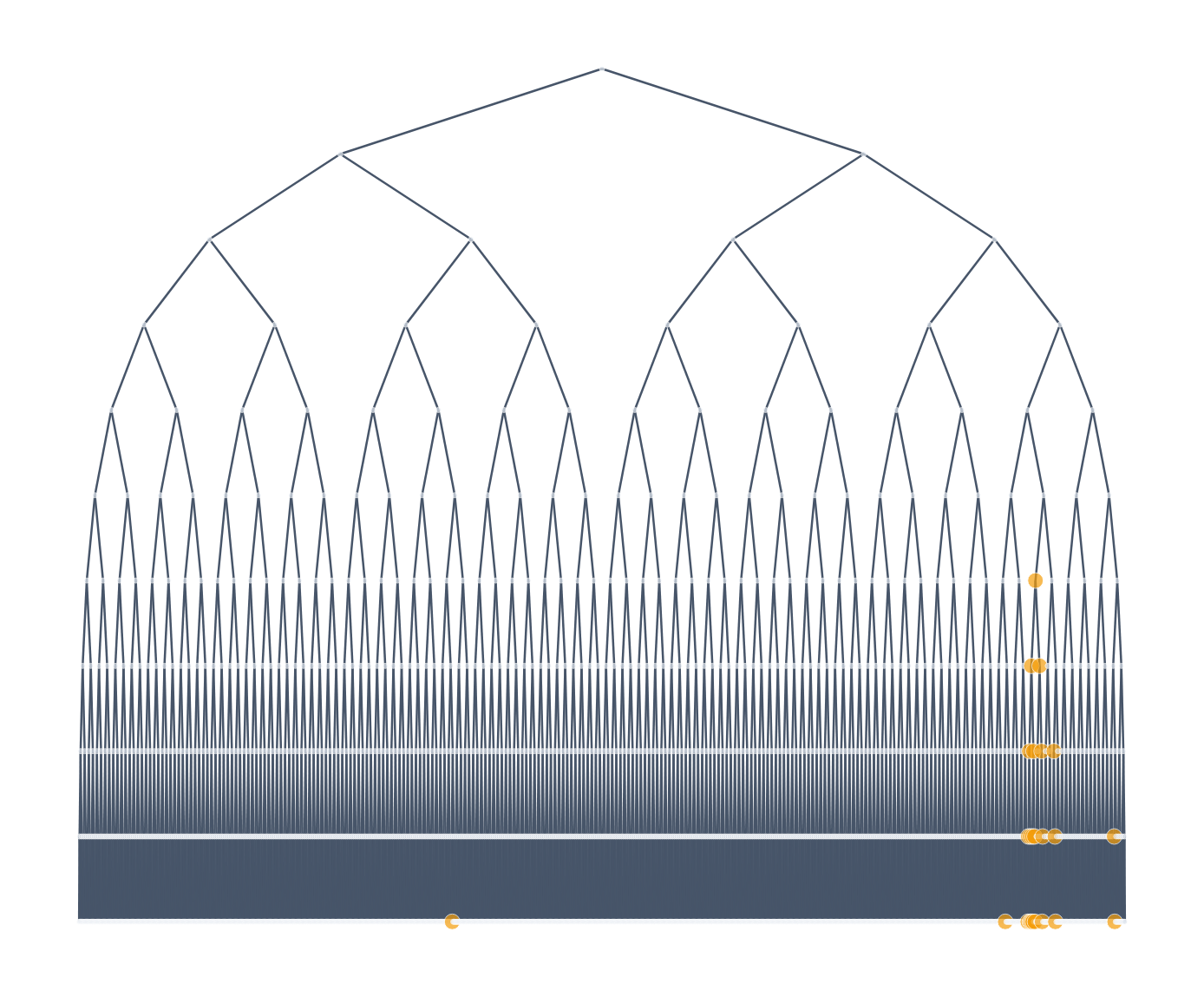} 
\end{tabular}
\caption{Interactive visualization tool features. \textit{Left:} Node selection highlights the root-to-node path with inline metadata annotations showing labels and keywords for each ancestor. \textit{Right:} Text search for ``physics'' highlights matching nodes in the tree.}
\label{fig:viz_tool_features}
\end{figure}

\begin{figure}[H]
    \centering
    \includegraphics[width=0.7\textwidth, angle=-90, origin=c]{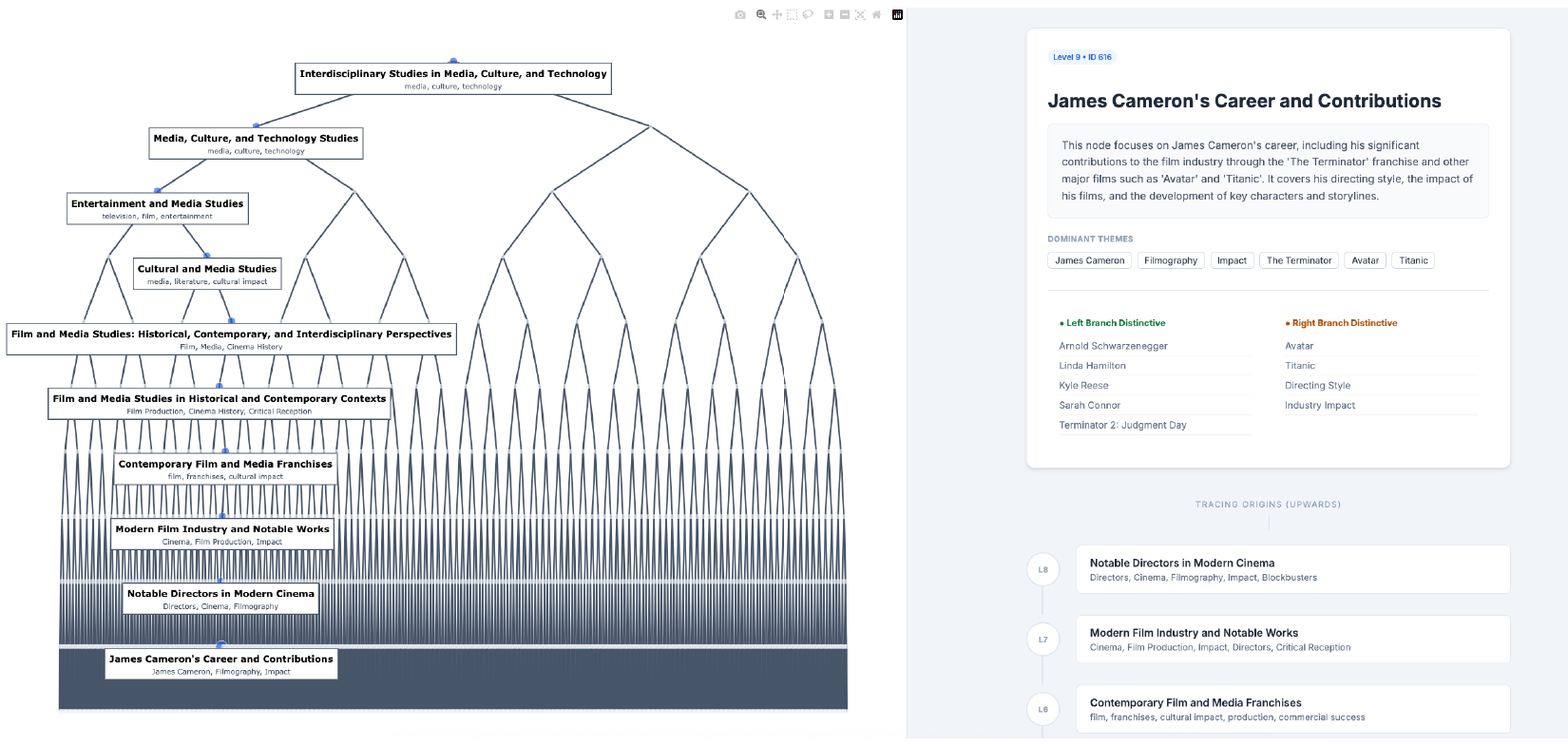}
    \vspace{-100pt}
    \caption{Visualization interface showing the interactive tree (left) and node metadata panel (right). The tree reveals a clear semantic refinement path (Entertainment $\rightarrow$ Contemporary Films $\rightarrow$ Notable Directors $\rightarrow$ James Cameron), while the metadata highlights the split into the \textit{Terminator} franchise (left subtree) versus other works such as \textit{Avatar} and \textit{Titanic} (right subtree).}
    \label{fig:james_cameron}
\end{figure}

\paragraph{Features} The tool supports interactive path highlighting: selecting a node highlights the full root-to-node path and shows each ancestor’s label and key keywords inline, making semantic refinement easy to track (see \cref{fig:viz_tool_features}). Navigation is streamlined via clickable ancestor breadcrumbs and left/right branch panels, each annotated with keywords for quick semantic orientation. A full-text search runs over node labels, summaries, and keywords, returning results grouped by tree level while simultaneously highlighting matches in the tree. The search box includes autocomplete over all node labels for direct jumps to relevant subtrees. For each node, we display top dominant themes and branch-distinctive keywords that summarize how the left and right children diverge (see \cref{fig:james_cameron}).

\begin{figure}[H]
\centering
\begin{tabular}{cc}
\includegraphics[width=0.40\textwidth, angle=-90, origin=c]{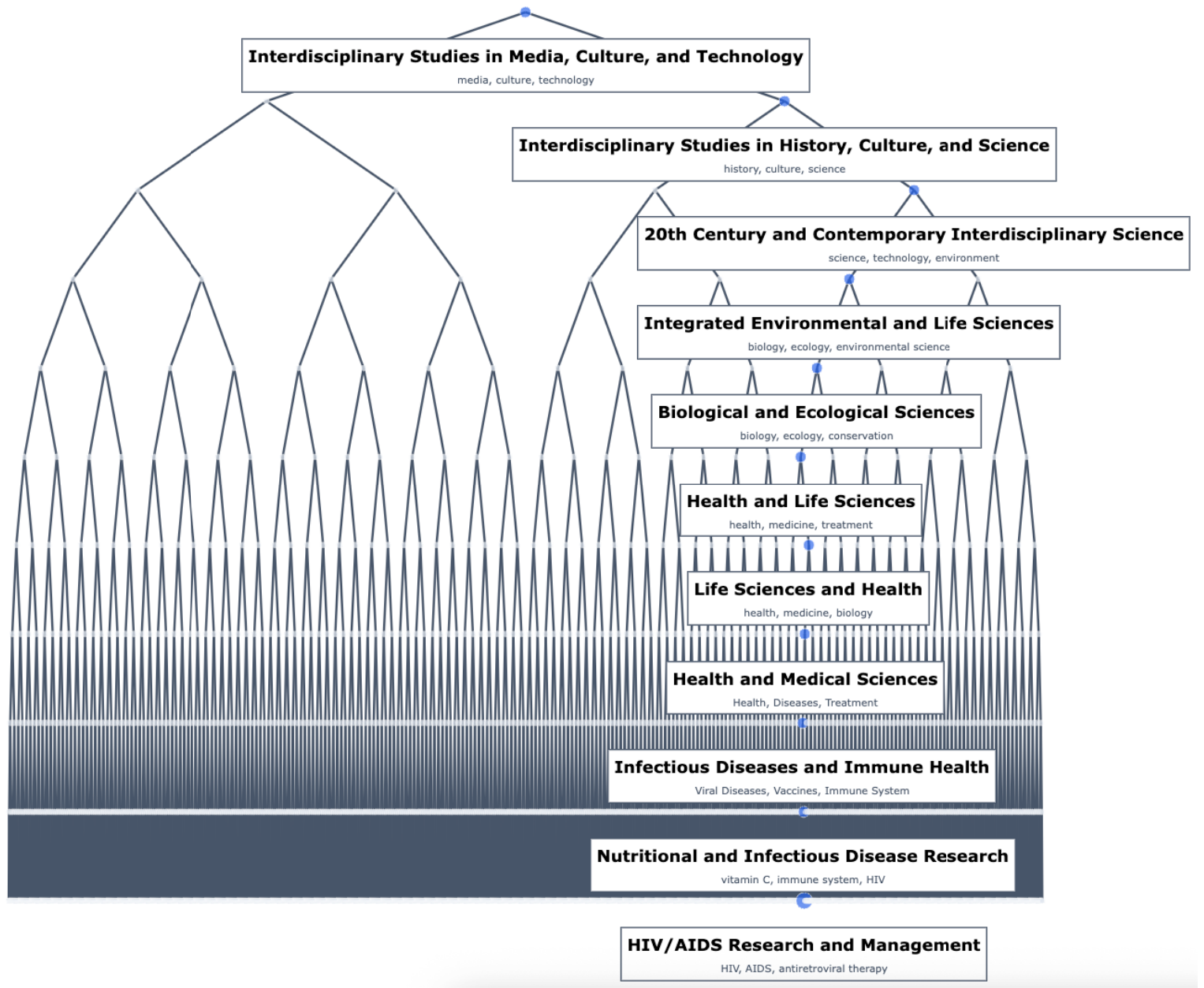} &
\includegraphics[width=0.40\textwidth, angle=-90, origin=c]{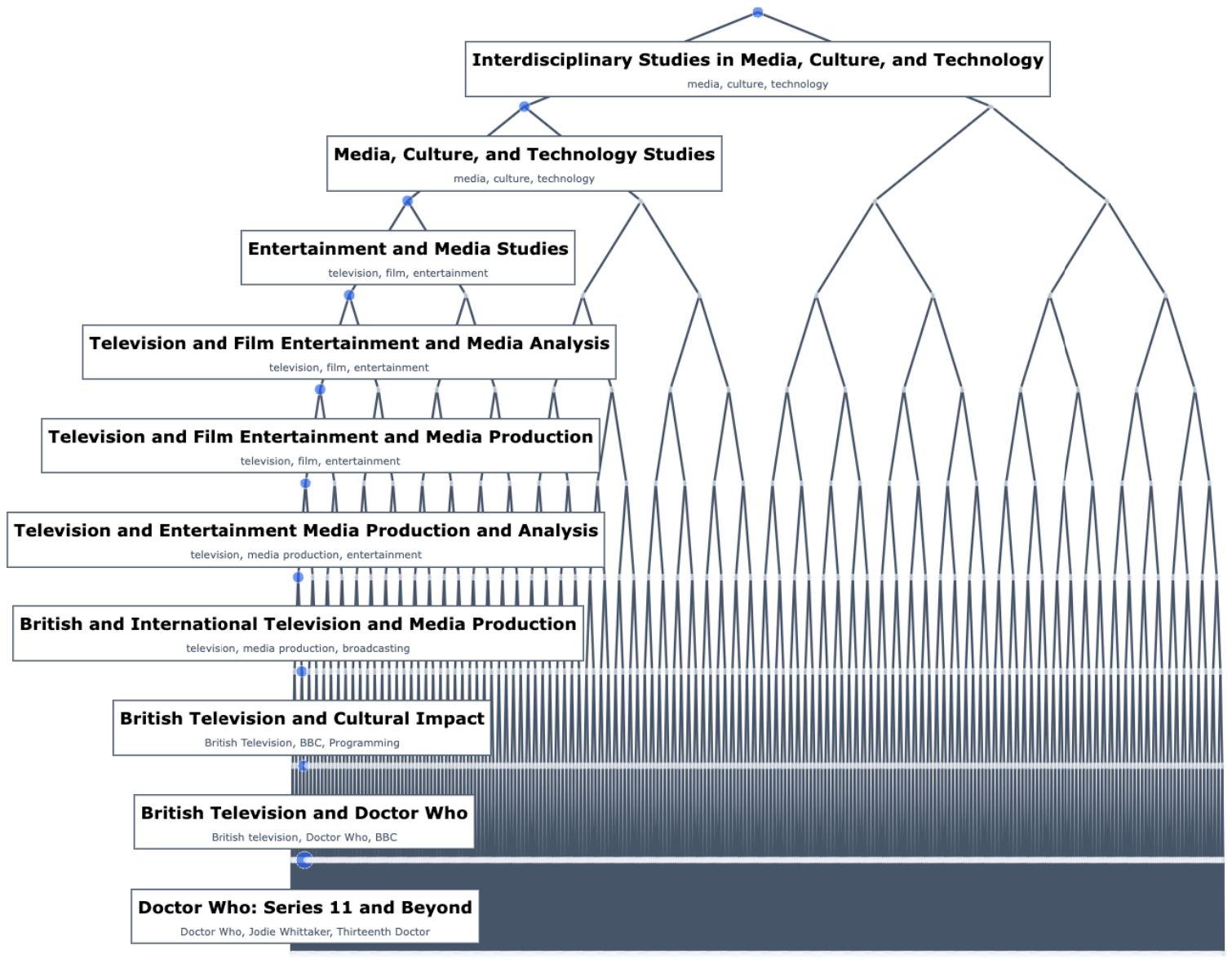} 
\end{tabular}
\caption{Examples of hierarchical document organization. \textit{Left}: Science $\rightarrow$ Biology $\rightarrow$ Health $\rightarrow$ Medicine $\rightarrow$ Infectious Diseases $\rightarrow$ HIV. \textit{Right}: Media $\rightarrow$ Entertainment $\rightarrow$ Television $\rightarrow$ British Television $\rightarrow$ Doctor Who $\rightarrow$ Doctor Who (Season 11).}
\label{fig:hierarchichal_org_examples}
\end{figure}

\paragraph{Case Studies: Hierarchical Organization}
The tree visualization provides an intuitive view of semantic refinement across levels, making it easy to understand how \ourmodel{} organizes the corpus into progressively finer concepts. For example, in \Cref{fig:james_cameron}, the left panel highlights a clear refinement path (Entertainment $\rightarrow$ Contemporary Films $\rightarrow$ Notable Directors $\rightarrow$ James Cameron), while the right panel reveals a meaningful split of the director’s filmography into the \textit{Terminator} franchise versus other major works (e.g., \textit{Avatar}, \textit{Titanic}). Additional examples in \cref{fig:hierarchichal_org_examples} further illustrate this refinement behavior: one branch narrows from Science $\rightarrow$ Biology $\rightarrow$ Health $\rightarrow$ Medicine to Infectious Diseases and ultimately HIV, while another specializes from Media $\rightarrow$ Entertainment $\rightarrow$ Television to British Television and finally to \textit{Doctor Who} (Season 11). These paths show that the tree captures both broad topical structure and highly specific subtopics at deeper levels.

\begin{figure}
    \centering
    \includegraphics[width=0.7\textwidth, angle=-90, origin=c]{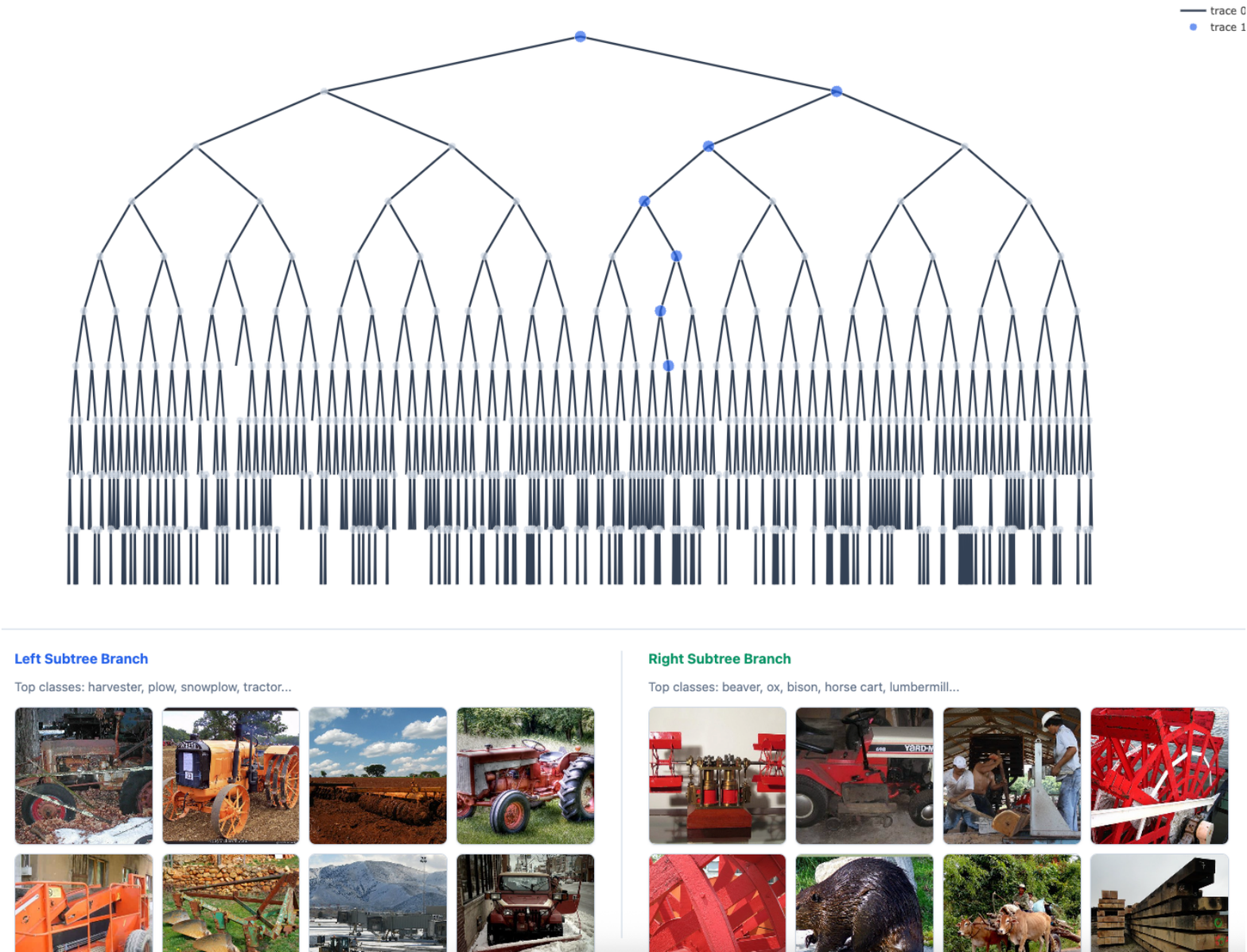}
    \vspace{-10pt}
    \caption{Example of a learned ImageNet1K hierarchy for image-to-image retrieval. The selected node partitions visually similar images into two coherent subclusters: modern agricultural machinery (left; e.g., harvesters, plows, snowplows) versus more primitive field-transport scenes (right; e.g., horse carts, oxen/bison pulling carts). This illustrates how the tree refines coarse visual similarity into finer-grained groupings.}
    \label{fig:harvester}
\end{figure}

\paragraph{ImageNet1K-Based Trees}
We next discuss \ourmodel{} trees trained on ImageNet1K for image-to-image retrieval, where the goal is to retrieve other images from the same class as a given query image. In this setting, the learned hierarchy primarily reflects visual similarity. For instance, \Cref{fig:harvester} shows a node that separates modern agricultural machinery (e.g., combine harvesters, plows, snowplows) in the left subtree from more primitive transport and field-work scenes in the right subtree (e.g., horse carts). The appearance of bison/oxen in the right subtree suggests that the split is driven by shared visual cues such as animals pulling carts in open fields, which can resemble harvester imagery at a coarse level while remaining semantically distinct, yielding a clean left/right partition.

\begin{figure}
    \centering
    \includegraphics[width=0.7\textwidth, angle=-90, origin=c]{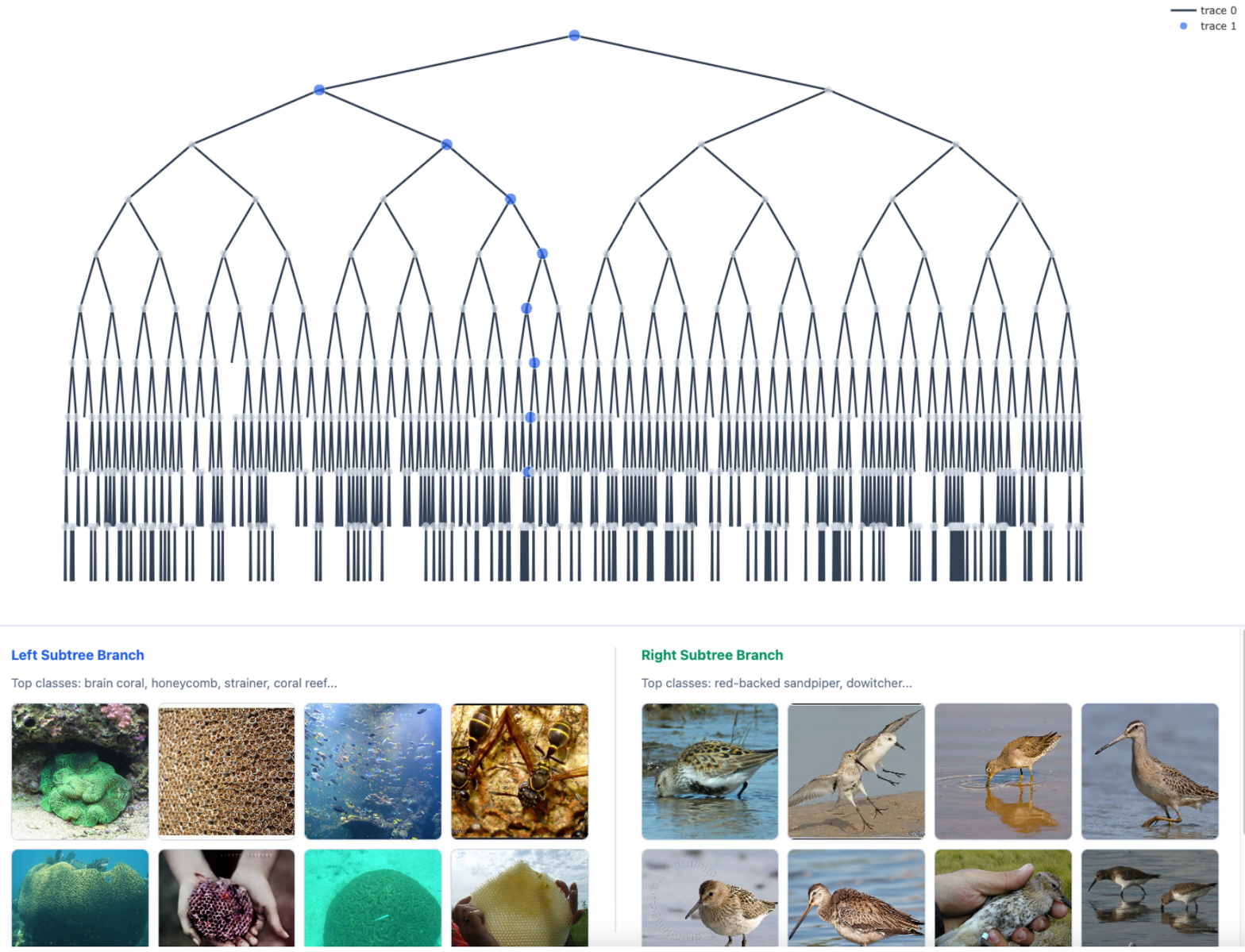}
    \vspace{-10pt}
    \caption{Example of visually driven hierarchy in an ImageNet1K ReTreever tree. The selected node groups images with similar high-frequency, porous or lattice-like textures. Its left subtree contains texture-dominated objects (coral leaves, honeycomb, strainers, brain coral), while its right subtree contains shoreline bird images (e.g., red-backed sandpiper, dowitcher). The split illustrates how the tree first captures broad visual similarity and then refines it into more specific visual concepts, consistent with an image-to-image retrieval objective.}
    \label{fig:honeycomb}
\end{figure}

Similarly, in \Cref{fig:honeycomb}, the selected node splits into two visually coherent subtrees. The left subtree groups coral leaves, honeycomb, strainers, and brain coral, which share strong lattice-like, porous, or repetitive textures. The right subtree contains birds such as the red-backed sandpiper and dowitcher; as waders, they are frequently photographed in shoreline scenes whose backgrounds (water, wet sand, ripples) can overlap visually with reef-like textures. At this level, the node captures a broad notion of ``near-water / high-frequency texture'' similarity, while its children separate these into texture-dominated objects (left) versus animal-centric shoreline scenes (right). This behavior is a direct consequence of the image-to-image retrieval objective, which prioritizes visual similarity over semantic relatedness.

\begin{figure}
    \centering
    \includegraphics[width=0.7\textwidth, angle=-90, origin=c]{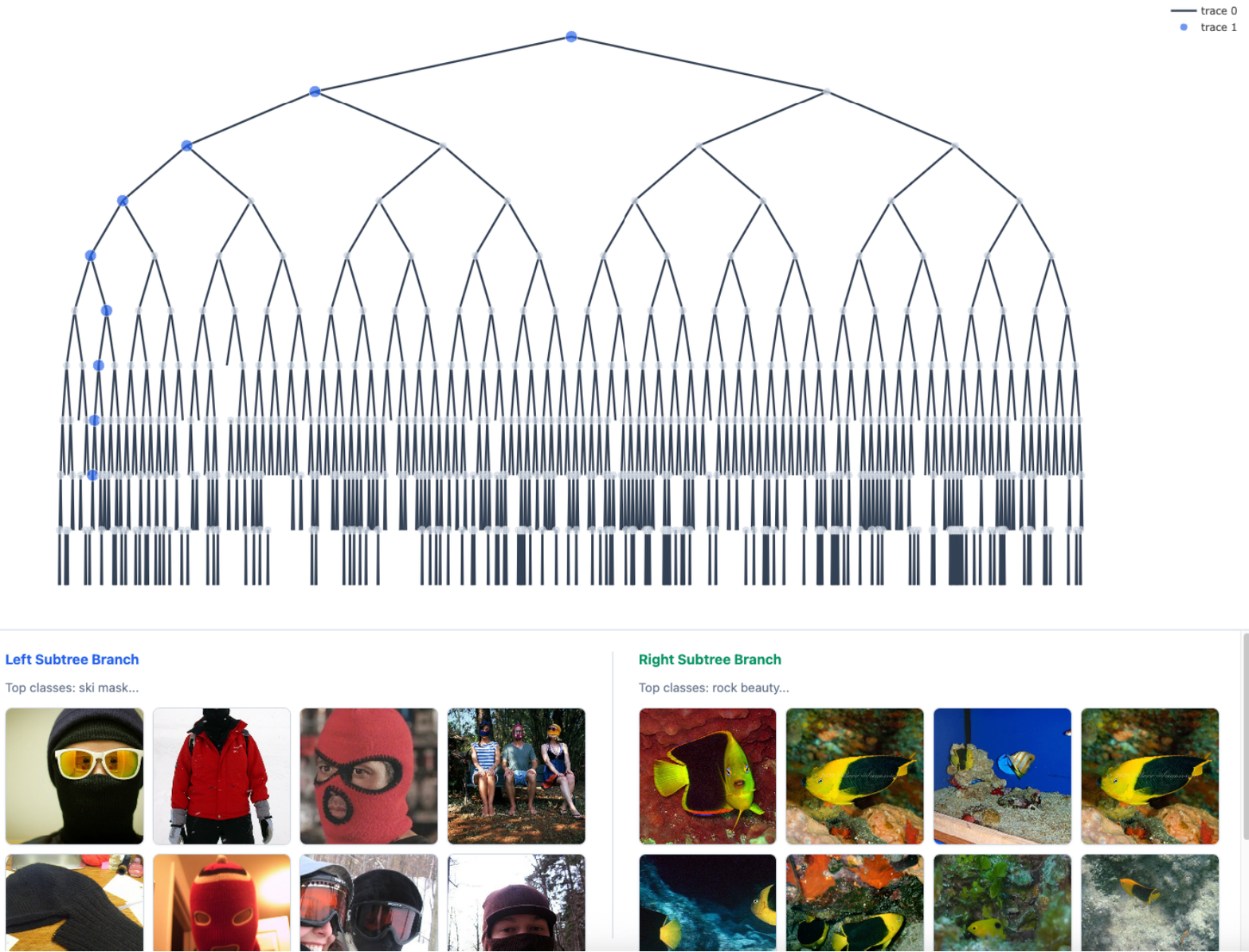}
    \vspace{-10pt}
    \caption{A visually coherent but semantically unexpected node: ski masks and rock beauty fish are grouped due to shared high-contrast, saturated color patterns (e.g., yellow/black/red), with the children splitting the two categories into separate subclusters.}
    \label{fig:skimask}
\end{figure}

\begin{figure}
    \centering
    \includegraphics[width=0.7\textwidth, angle=-90, origin=c]{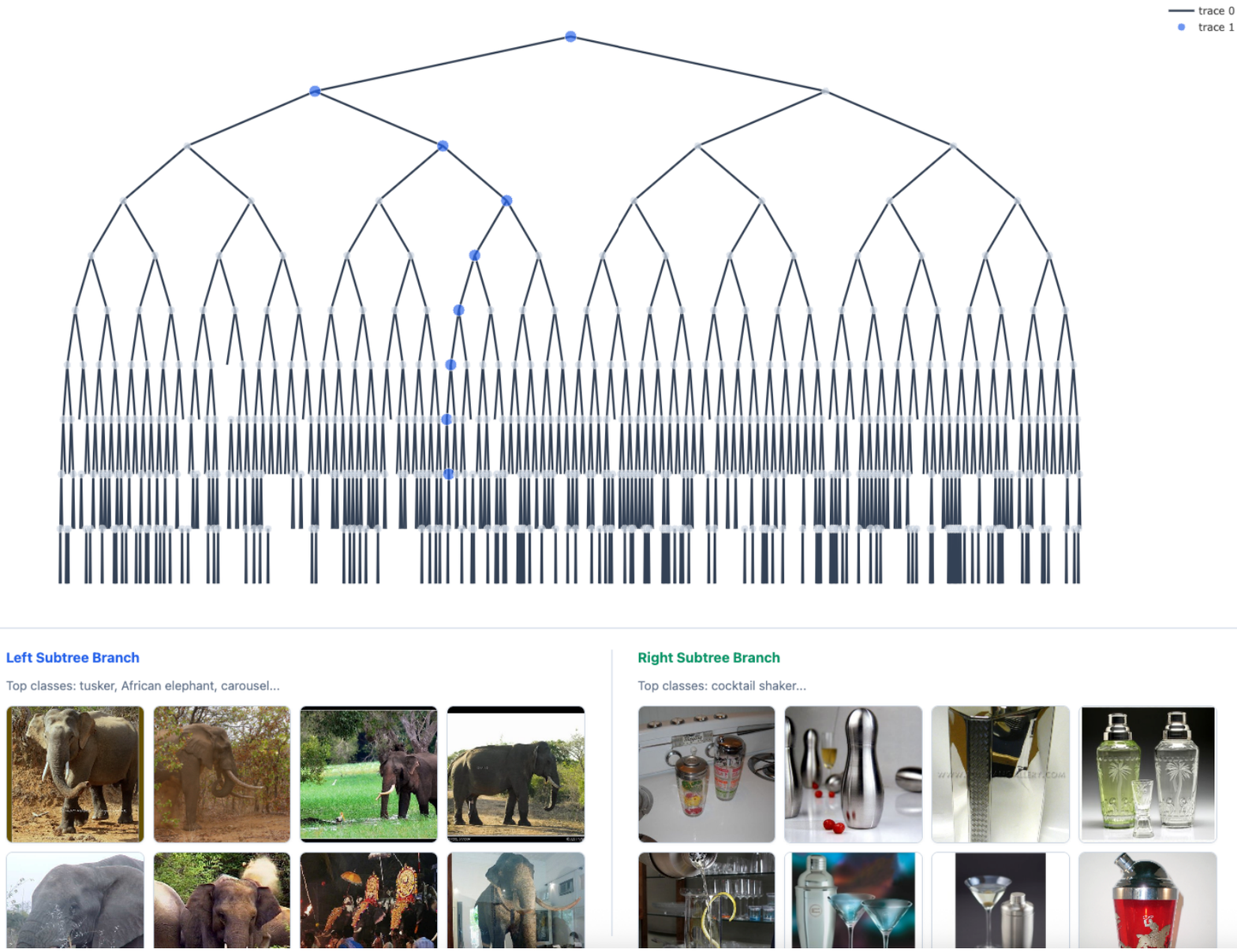}
    \vspace{-10pt}
    \caption{A shape-driven grouping: elephants and cocktail shakers co-occur at an internal node, likely reflecting a shared elongated curved silhouette (trunk-like geometry), before the children refine the cluster into elephants (left) versus cocktail shakers (right).}
    \label{fig:elephants}
\end{figure}

Finally, this kind of visually driven routing can also yield groupings that sound nonsensical textually, but are coherent in pixel-space. For example, in \Cref{fig:skimask}, a node clusters ski masks with rock beauty fish. While semantically unrelated, both categories frequently exhibit high-contrast, saturated color patterns (notably yellow, black, and red), and the node appears to capture this shared visual motif; the children then separate the two classes. Similarly, \Cref{fig:elephants} shows a node grouping elephants with cocktail shakers, which becomes intuitive on inspection: the elongated, curved silhouette of a shaker can resemble an elephant’s trunk. The subsequent split refines this coarse visual similarity into distinct subclusters, with elephants in the left subtree and cocktail shakers in the right.

We hope readers find our visualization tool useful and enjoyable to explore, and that it highlights how trees learned by \ourmodel{} can be readily converted into artifacts for direct visual inspection and analysis.

\section{Hyperparameters}
\label{app:hyperparameters}

This appendix details the hyperparameters used for training \ourmodel{} models across different modalities and datasets.

\subsection{Common Training Hyperparameters}

Table~\ref{tab:common_hyperparams} lists hyperparameters shared across all \ourmodel{} variants.

\begin{table}[h]
\centering
\caption{Common training hyperparameters for all ReTreever models.}
\label{tab:common_hyperparams}
\begin{tabular}{lc}
\toprule
\textbf{Hyperparameter} & \textbf{Value} \\
\midrule
Optimizer & AdamW \\
Learning rate & 4e-4 \\
Weight decay & 0.01 \\
Warmup steps & 10,000 \\
LR Scheduler & Linear \\
Total training steps & 200,000 \\
Batch size & 64 \\
Gradient clipping & 1.0 \\
Tree depth ($D$) & 10 \\
Maximum leaves & 1,024 \\
Loss function & InfoNCE \cref{eq:constrative_loss} \\
Similarity metric & Negative TVD \\
\bottomrule
\end{tabular}
\end{table}

\subsection{Architecture Hyperparameters}

Table~\ref{tab:arch_hyperparams} describes the architecture-specific hyperparameters for the tree structure and split functions.

\begin{table}[h]
\centering
\caption{ReTreever architecture hyperparameters.}
\label{tab:arch_hyperparams}
\begin{tabular}{lc}
\toprule
\textbf{Hyperparameter} & \textbf{Value} \\
\midrule
\multicolumn{2}{l}{\textit{Split Function}} \\
Split function type & Cross-attention \\
Number of attention heads & 16 \\
Attention head dimension ($d_k$) & 64 \\
Total split function dimension & 1024 \\
Embeddings per level ($n_l$) & 8 \\
Level embedding dimension & 1,024 \\
\midrule
\multicolumn{2}{l}{\textit{Tree Propagation}} \\
Propagation method & product-propagation \\
Aggregation function & Per-node linear projection + mean \\
\midrule
\multicolumn{2}{l}{\textit{Depth Scheduling}} \\
Constant depth & Level 10 only \\
Stochastic depth & Levels 1--10, $p(\ell)\propto \ell^2$ \\
\bottomrule
\end{tabular}
\end{table}

\subsection{Encoder-Specific Hyperparameters}

Table~\ref{tab:encoder_hyperparams} details encoder configurations across all modalities. For text retrieval we use DistilBERT (trained on MS MARCO) as our primary encoder and BGE-large as a secondary comparison. Image retrieval experiments use three vision encoders on ImageNet-1K, and audio retrieval uses AST on VoxCeleb2.
\begin{table}
\caption{Encoder hyperparameters across modalities.}
\label{tab:encoder_hyperparams}
\resizebox{\textwidth}{!}{%
\begin{tabular}{l|cc|ccc|c}
\toprule
& \multicolumn{2}{c|}{\textbf{Text}} & \multicolumn{3}{c|}{\textbf{Image}} & \textbf{Audio} \\
\textbf{Hyperparameter} & \textbf{DistilBERT} & \textbf{BGE} & \textbf{DinoV2} & \textbf{ResNet50} & \textbf{CLIP-ViT-L} & \textbf{AST} \\
\midrule
Base encoder & distilbert-base & bge-large-en & dinov2-large & resnet50 & clip-vit-large & ast-audioset \\
 & (MS MARCO) & v1.5 & & & patch14 & \\
Number of parameters & 66M & 335M & 407M & 26M & 304M & 86M \\
Encoder dimension & 768 & 1,024 & 1,024 & 2,048 & 768 & 768 \\
Max sequence length & 512 tokens & 512 tokens & -- & -- & -- & -- \\
Image resolution & -- & -- & 224×224 & 224×224 & 224×224 & -- \\
Audio sampling rate & -- & -- & -- & -- & -- & 16 kHz \\
Audio duration & -- & -- & -- & -- & -- & 10 sec \\
\bottomrule
\end{tabular}%
}
\end{table}

\subsection{Hardware}

All models were trained on NVIDIA H100 80GB GPUs.

\clearpage

\newpage

\end{document}